%% file: tplp.tex
\documentclass{tlp}

\makeatletter
\let\O@argtabularcr\@argtabularcr
\def\O@xtabularcr{\@ifnextchar[\O@argtabularcr{\ifnum 0=`{\fi}\cr}}
\let\O@tabacol\@tabacol
\let\O@tabclassiv\@tabclassiv
\let\O@tabclassz\@tabclassz
\let\O@tabarray\@tabarray
\def\author@tabular{\authorsize\def\@halignto{}\@authortable}
\let\endauthor@tabular=\endtabular
\def\author@tabcrone{{\ifnum0=`}\fi\O@xtabularcr\affilsize\itshape
 \let\\=\author@tabcrtwo\ignorespaces}
\def\author@tabcrtwo{{\ifnum0=`}\fi\O@xtabularcr[-3\p@]\affilsize\itshape
 \let\\=\author@tabcrtwo\ignorespaces}
\def\@authortable{\leavevmode \hbox \bgroup $\let\@acol\O@tabacol
 \let\@classz\O@tabclassz \let\@classiv\O@tabclassiv
 \let\\=\author@tabcrone \ignorespaces \O@tabarray}
\makeatother

\usepackage{mathptmx}

\usepackage{url}
\usepackage[breaklinks]{hyperref}
\usepackage{
  booktabs,
  siunitx,
  csquotes,
  todonotes,
  enumitem,
  amsmath
}

\usepackage{dcolumn}
\usepackage{booktabs}
\usepackage{tikz}
\usepackage{pgfplots}
\pgfplotsset{compat=newest}
\usetikzlibrary{patterns}

\newcolumntype{M}[1]{D{.}{.}{1.#1}}

\renewcommand{\figurename}{Figure}

\colorlet{curry}{orange!70!white}
\colorlet{probLog}{violet!70!white}
\colorlet{webPPL}{blue!70!black}

\makeatletter
\@ifundefined{lhs2tex.lhs2tex.sty.read}%
  {\@namedef{lhs2tex.lhs2tex.sty.read}{}%
   \newcommand\SkipToFmtEnd{}%
   \newcommand\EndFmtInput{}%
   \long\def\SkipToFmtEnd#1\EndFmtInput{}%
  }\SkipToFmtEnd

\newcommand\ReadOnlyOnce[1]{\@ifundefined{#1}{\@namedef{#1}{}}\SkipToFmtEnd}
\usepackage{amstext}
\usepackage{amssymb}
\usepackage{stmaryrd}
\DeclareFontFamily{OT1}{cmtex}{}
\DeclareFontShape{OT1}{cmtex}{m}{n}
  {<5><6><7><8>cmtex8
   <9>cmtex9
   <10><10.95><12><14.4><17.28><20.74><24.88>cmtex10}{}
\DeclareFontShape{OT1}{cmtex}{m}{it}
  {<-> ssub * cmtt/m/it}{}

\DeclareFontShape{OT1}{cmtt}{bx}{n}
  {<5><6><7><8>cmtt8
   <9>cmbtt9
   <10><10.95><12><14.4><17.28><20.74><24.88>cmbtt10}{}
\DeclareFontShape{OT1}{cmtex}{bx}{n}
  {<-> ssub * cmtt/bx/n}{}

\newcommand{\Conid}[1]{\mathit{#1}}
\newcommand{\Varid}[1]{\mathit{#1}}
\newcommand{\anonymous}{\kern0.06em \vbox{\hrule\@width.5em}}
\newcommand{\plus}{\mathbin{+\!\!\!+}}

\renewcommand{\geq}{\geqslant}
\usepackage{polytable}

\@ifundefined{mathindent}%
  {\newdimen\mathindent\mathindent\leftmargini}%
  {}%

\def\resethooks{%
  \global\let\SaveRestoreHook\empty
  \global\let\ColumnHook\empty}
\newcommand*{\savecolumns}[1][default]%
  {\g@addto@macro\SaveRestoreHook{\savecolumns[#1]}}
\newcommand*{\restorecolumns}[1][default]%
  {\g@addto@macro\SaveRestoreHook{\restorecolumns[#1]}}
\newcommand*{\aligncolumn}[2]%
  {\g@addto@macro\ColumnHook{\column{#1}{#2}}}

\resethooks

\newcommand{\onelinecommentchars}{\quad-{}- }
\newcommand{\commentbeginchars}{\enskip\{-}
\newcommand{\commentendchars}{-\}\enskip}

\newcommand{\visiblecomments}{%
  \let\onelinecomment=\onelinecommentchars
  \let\commentbegin=\commentbeginchars
  \let\commentend=\commentendchars}

\newcommand{\invisiblecomments}{%
  \let\onelinecomment=\empty
  \let\commentbegin=\empty
  \let\commentend=\empty}

\visiblecomments

\newlength{\blanklineskip}
\newcommand{\hsindent}[1]{\quad}
\let\hspre\empty
\let\hspost\empty

\EndFmtInput
\makeatother
\ReadOnlyOnce{polycode.fmt}%
\makeatletter

\newcommand{\hsnewpar}[1]%
  {{\parskip=0pt\parindent=0pt\par\vskip #1\noindent}}

\newcommand{\hscodestyle}{}

\newcommand{\sethscode}[1]%
  {\expandafter\let\expandafter\hscode\csname #1\endcsname
   \expandafter\let\expandafter\endhscode\csname end#1\endcsname}

  {\par\noindent
   \advance\leftskip\mathindent
   \hscodestyle
   \let\\=\@normalcr
   \let\hspre\(\let\hspost\)%
   \pboxed}%
  {\endpboxed\)%
   \par\noindent
   \ignorespacesafterend}

  {\hsnewpar\abovedisplayskip
   \advance\leftskip\mathindent
   \hscodestyle
   \let\hspre\(\let\hspost\)%
   \pboxed}%
  {\endpboxed%
   \hsnewpar\belowdisplayskip
   \ignorespacesafterend}

  {\hsnewpar\abovedisplayskip
   \advance\leftskip\mathindent
   \hscodestyle
   \let\\=\@normalcr
   \(\pboxed}%
  {\endpboxed\)%
   \hsnewpar\belowdisplayskip
   \ignorespacesafterend}

\newcommand{\plainhs}{\sethscode{plainhscode}}

\plainhs

  {\hsnewpar\abovedisplayskip
   \advance\leftskip\mathindent
   \hscodestyle
   \let\\=\@normalcr
   \(\parray}%
  {\endparray\)%
   \hsnewpar\belowdisplayskip
   \ignorespacesafterend}

  {\parray}{\endparray}

  {\(\parray}{\endparray\)}

\def\codeframewidth{\arrayrulewidth}
\RequirePackage{calc}

  {\parskip=\abovedisplayskip\par\noindent
   \hscodestyle
   \arrayrulewidth=\codeframewidth
   \tabular{@{}|p{\linewidth-2\arraycolsep-2\arrayrulewidth-2pt}|@{}}%
   \hline\framedhslinecorrect\\{-1.5ex}%
   \let\endoflinesave=\\
   \let\\=\@normalcr
   \(\pboxed}%
  {\endpboxed\)%
   \framedhslinecorrect\endoflinesave{.5ex}\hline
   \endtabular
   \parskip=\belowdisplayskip\par\noindent
   \ignorespacesafterend}

\newcommand{\framedhslinecorrect}[2]%
  {#1[#2]}

  {\(\def\column##1##2{}%
   \let\>\undefined\let\<\undefined\let\\\undefined
   \newcommand\>[1][]{}\newcommand\<[1][]{}\newcommand\\[1][]{}%
   \def\fromto##1##2##3{##3}%
   }{\) }%

  {\let\orighscode=\hscode
   \let\origendhscode=\endhscode
   \def\endhscode{\def\hscode{\endgroup\def\@currenvir{hscode}\\}\begingroup}
   \orighscode\def\hscode{\endgroup\def\@currenvir{hscode}}}%
  {\origendhscode
   \global\let\hscode=\orighscode
   \global\let\endhscode=\origendhscode}%

\makeatother
\EndFmtInput
\ReadOnlyOnce{forall.fmt}%
\makeatletter

\let\HaskellResetHook\empty
\newcommand*{\AtHaskellReset}[1]{%
  \g@addto@macro\HaskellResetHook{#1}}
\newcommand*{\HaskellReset}{\HaskellResetHook}

\newcommand\hsforall{\global\let\hsdot=\hsperiodonce}
\newcommand*\hsperiodonce[2]{#2\global\let\hsdot=\hscompose}
\newcommand*\hscompose[2]{#1}

\AtHaskellReset{\global\let\hsdot=\hscompose}

\HaskellReset

\makeatother
\EndFmtInput

\def\commentbegin{\quad\{\ }
\def\commentend{\}}
  \title[Theory and Practice of Logic Programming]
        {Implementing a Library for Probabilistic Programming using Non-strict Non-determinism\footnote{This is an extended version of a paper presented at the \emph{International Symposium on Practical Aspects of Declarative Languages (PADL 2018)}, invited as a rapid communication in TPLP. The authors acknowledge the assistance of the conference program chairs Nicola Leone and Kevin Hamlen.}}

  \author[S. Dylus \and J. Christiansen \and F. Teegen]
         {SANDRA DYLUS\\
         University of Kiel\\
         \email{sad@informatik.uni-kiel.de}
         \and
         JAN CHRISTIANSEN\\
         Flensburg University of Applied Sciences\\
         \email{jan.christiansen@hs-flensburg.de}
         \and
         FINN TEEGEN\\
         University of Kiel\\
         \email{fte@informatik.uni-kiel.de}}

\jdate{March 2003}
\pubyear{2003}
\pagerange{\pageref{firstpage}--\pageref{lastpage}}
\doi{S1471068401001193}

\begin{document}

\label{firstpage}

\maketitle

\begin{abstract}
This paper presents \emph{PFLP}, a library for probabilistic programming in the functional logic programming language Curry.
It demonstrates how the concepts of a functional logic programming language support the implementation of a library for probabilistic programming.
In fact, the paradigms of functional logic and probabilistic programming are closely connected.
That is, language characteristics from one area exist in the other and vice versa.
For example, the concepts of non-deterministic choice and call-time choice as known from functional logic programming are related to and coincide with stochastic memoization and probabilistic choice in probabilistic programming, respectively.
We will further see that an implementation based on the concepts of functional logic programming can have benefits with respect to performance compared to a standard list-based implementation and can even compete with full-blown probabilistic programming languages, which we illustrate by several benchmarks.
\textit{Under consideration in Theory and Practice of Logic Programming (TPLP).}
\end{abstract}



\section{Introduction}

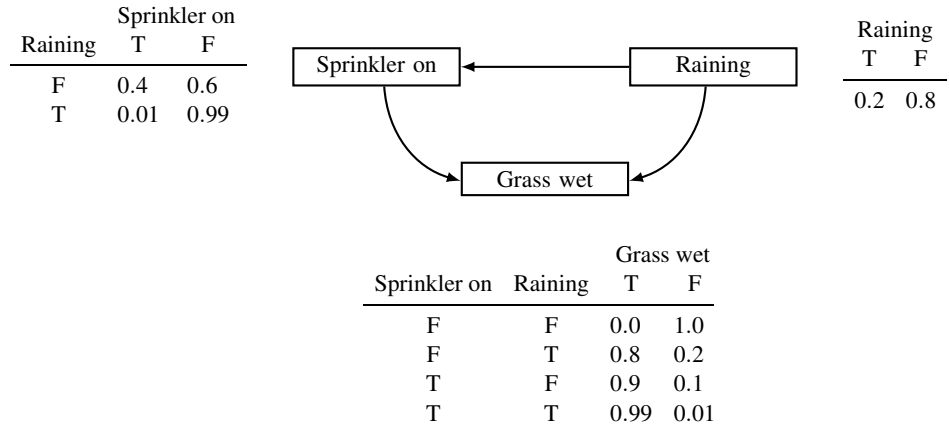
\begin{figure}
\begin{tikzpicture}[
  node distance=1cm and 0cm,
  mynode/.style={draw,rectangle,thick,text width=2cm,align=center},
  scale=1.0,
  every node/.style={scale=1.0}
]
\node[mynode] (sp) {Sprinkler on};
\node[mynode,below right=of sp] (gw) {Grass wet};
\node[mynode,above right=of gw] (ra) {Raining};
\path (ra) edge[-latex,thick] (sp)
(sp) edge[-latex,thick,bend right=35] (gw)
(gw) edge[latex-,thick,bend right=35] (ra);
\node[left=0.5cm of sp]
{
\begin{tabular}{cM{2}M{2}}
& \multicolumn{2}{c}{Sprinkler on} \\
Raining & \multicolumn{1}{c}{T} & \multicolumn{1}{c}{F} \\
\cmidrule(r){1-2}\cmidrule(l){2-3}
F & 0.4 & 0.6 \\
T & 0.01 & 0.99
\end{tabular}
};
\node[right=0.5cm of ra]
{
\begin{tabular}{M{1}M{1}}
\multicolumn{2}{c}{Raining} \\
\multicolumn{1}{c}{T} & \multicolumn{1}{c}{F} \\
\cmidrule{1-2}
0.2 & 0.8
\end{tabular}
};
\node[below=0.5cm of gw]
{
\begin{tabular}{ccM{2}M{2}}
& & \multicolumn{2}{c}{Grass wet} \\
Sprinkler on & Raining & \multicolumn{1}{c}{T} & \multicolumn{1}{c}{F} \\
\cmidrule(r){1-4}
F & F & 0.0 & 1.0 \\
F & T & 0.8 & 0.2 \\
T & F & 0.9 & 0.1 \\
T & T & 0.99 & 0.01
\end{tabular}
};
\end{tikzpicture}
\caption{A simple Bayesian network}
\label{fig:bayes}
\end{figure}

The probabilistic programming paradigm allows the succinct definition of probabilistic processes and other applications based on probability distributions, for example, Bayesian networks as used in machine learning.
A Bayesian network is a visual, graph-based representation for a set of random variables and their dependencies.
One of the \emph{hello world}-examples of Bayesian networks is the influence of rain and a sprinkler on wet grass.
\autoref{fig:bayes} shows an instance of this example, the concrete probabilities differ between publications.
A node in the graph represents a random variable, a directed edge between two nodes represents a conditional dependency.
Each node is annotated with a probability function represented as a table.
The input values are on the left side of the table and the right side of the table describes the possible output and the corresponding probability.
The input values of the function correspond to the incoming edges of that node.
For example, the node for sprinkler depends on rain, thus, the sprinkler node has an incoming edge that originates from the rain node.
The input parameter rain appears directly in the table that describes the probability function for sprinkler.
For the example in \autoref{fig:bayes} the interpretation of the graph reads as follows: it rains with a probability of \SI{20}{\percent}; depending on the rain, the probability for an activated sprinkler is \SI{40}{\percent} and \SI{1}{\percent}, respectively; depending on both these factors, the grass can be observed as wet with a probability of \SI{0}{\percent}, \SI{80}{\percent}, \SI{90}{\percent} or \SI{99}{\percent}.
The network can answer the following exemplary questions.

\begin{itemize}
\item What is the probability that it is raining?
\item What is the probability that the grass is wet, given that it is raining?
\item What is the probability that the sprinkler is on, given that the grass is wet?
\end{itemize}

The general idea of probabilistic programming has been quite successful.
There are a variety of probabilistic programming languages supporting all kinds of programming paradigms.
For example, the programming languages Church~\cite{goodman2008church}
and Anglican~\cite{wood2014new} are based on the functional
programming language Scheme, ProbLog~\cite{kimmig2011implementation} is an
extension of the logic programming language Prolog, Probabilistic
C~\cite{paige2014compilation} is based on the imperative language C,
and WebPPL~\cite{goodman2014design}, the successor of Church, is embedded in a functional subset of JavaScript.
Besides full-blown languages there are also embedded domain-specific languages that implement probabilistic programming as a library.
For example, FACTORIE~\cite{mccallum2009factorie} is a library for the hybrid programming language Scala, and Erwig and Kollmansberger \cite{erwig2006functional} present a library for the functional programming language Haskell.
We recommend the survey by Gordon et al. \cite{gordon2014probabilistic} about the current state of probabilistic programming for further information.

This paper presents \emph{PFLP}, a library providing a domain-specific language for probabilistic programming in the functional logic programming language Curry~\cite{antoy2010functional}.
PFLP makes heavy use of functional logic programming concepts and shows that this paradigm is well-suited for implementing a library for probabilistic programming.
In fact, there is a close connection between probabilistic programming and functional logic programming.
For example, non-deterministic choice and probabilistic choice are similar concepts.
Furthermore, the concept of call-time choice as known from functional logic programming coincides with (stochastic) memoization \cite{deraedt2013probabilistic} in the area of probabilistic programming.
We are not the first to observe this close connection between functional logic programming and probabilistic programming.
For example,~Fischer et al. \cite{fischer2009purely} present a library for modeling functional logic programs in the functional language Haskell.
As they state, by extending their approach to weighted non-determinism we can model a probabilistic programming language.

Besides a lightweight implementation of a library for probabilistic programming in a functional logic programming language, this paper makes the following contributions.
\begin{itemize}
\item We investigate the interplay of probabilistic programming with the features of a functional logic programming language.
For example, we show how call-time choice and non-determinism interplay with probabilistic choice.
\item We discuss how we utilize functional logic features to improve the implementation of probabilistic combinators.
\item We present an implementation of probability distributions using non-determinism in combination with non-strict probabilistic combinators that is more efficient than an implementation using lists.
\item We illustrate that the combination of non-determinism and non-strict\-ness with respect to distributions has to be handled with care.
More precisely, it is important to enforce a certain degree of
strictness in order to guarantee correct results.
\item In contrast to the conference version of the paper \cite{dylus2018probabilistic} we discuss the usage of partial functions in combination with library functions in more detail, reason about laws for two operations of the library, and present performance comparisons between our library, ProbLog and WebPPL.
\item Finally, this paper aims at fostering the exchange between the community of probabilistic programming and of functional logic programming.
That is, while the connection exists for a long time, there has not been much exchange between the communities.
We would like to take this paper as a starting point to bring these paradigms closer together.
Thus, this paper introduces the concepts of both, the functional logic and probabilistic programming, paradigms.
\end{itemize}

Please note that the current state of our library cannot compete against full-blown probabilistic languages or mature libraries for probabilistic programming in terms of features, e.g., the library does not provide any sampling mechanisms.
Nevertheless, the library is a good showcase for languages with built-in non-determinism, because the functional logic approach can be superior to the functional approach using lists.
Furthermore, we want to emphasize that this paper uses non-determinism as an implementation technique to develop a library for probabilistic programming.
That is, we are not mainly concerned with the interaction of non-determinism and probabilities as, for example, discussed in the work of Varacca and Winskel \cite{varacca2006distributing} and multiple others.
The library we develop in this paper does not combine both effects, but provides combinators for probabilistic programming by leveraging Curry's built-in non-strict non-determinism.

\section{Library Basics}
\label{sec:idea}

In this section we discuss the core of the PFLP library\footnote{We provide the code for the library at \url{https://github.com/finnteegen/pflp}.}.
The implementation is based on a Haskell library for probabilistic
programming presented by Erwig and Kollmansberger \cite{erwig2006functional}.
We will not present the whole PFLP library, but only core functions.
The paper at hand is a literate Curry file. We use the Curry compiler KiCS2\footnote{We use version \text{\tt 0\char46{}6\char46{}0} of KiCS2 and the source is found at \url{https://www-ps.informatik.uni-kiel.de/kics2/}.} by Bra\ss{}el et al. \cite{brassel2011kics2} for all code examples.

\subsection{Modeling Distributions}

One key ingredient of probabilistic programming is the definition of distributions.
A distribution consists of pairs of elementary events and their probability.
We model probabilities as \ensuremath{\Conid{Float}} and distributions as a combination of an elementary event and the corresponding probability.\footnote{The polymorph data type \ensuremath{\Conid{Dist}} is parameterized over a type variable \ensuremath{\Varid{a}}. It has a single constructor also named \ensuremath{\Conid{Dist}} that is of type \ensuremath{\Varid{a}\to \Conid{Probability}\to \Conid{Dist}\;\Varid{a}}. The constructor \ensuremath{\Conid{Dist}} in Curry corresponds to a binary functor in Prolog.}
\begin{hscode}\SaveRestoreHook
\column{B}{@{}>{\hspre}l<{\hspost}@{}}%
\column{3}{@{}>{\hspre}l<{\hspost}@{}}%
\column{E}{@{}>{\hspre}l<{\hspost}@{}}%
\>[3]{}\mathbf{type}\;\Conid{Probability}\mathrel{=}\Conid{Float}{}\<[E]%
\\
\>[3]{}\mathbf{data}\;\Conid{Dist}\;\Varid{a}\mathrel{=}\Conid{Dist}\;\Varid{a}\;\Conid{Probability}{}\<[E]%
\ColumnHook
\end{hscode}\resethooks

In a functional language like Haskell, the canonical way to define distributions uses lists.
Here, we use Curry's built-in non-determinism as an alternative for lists to model distributions with more than one event-probability pair.
As an example, we define a fair coin, where \ensuremath{\Conid{True}} represents heads and \ensuremath{\Conid{False}} represents tails, as follows.\footnote{Here and in the following we write probabilities as fractions for readability.}
\begin{hscode}\SaveRestoreHook
\column{B}{@{}>{\hspre}l<{\hspost}@{}}%
\column{3}{@{}>{\hspre}l<{\hspost}@{}}%
\column{E}{@{}>{\hspre}l<{\hspost}@{}}%
\>[3]{}\Varid{coin}\mathbin{::}\Conid{Dist}\;\Conid{Bool}{}\<[E]%
\\
\>[3]{}\Varid{coin}\mathrel{=}(\Conid{Dist}\;\Conid{True}\;\frac{\mathrm{1}}{\mathrm{2}})\mathbin{?}(\Conid{Dist}\;\Conid{False}\;\frac{\mathrm{1}}{\mathrm{2}}){}\<[E]%
\ColumnHook
\end{hscode}\resethooks

In Curry the \ensuremath{(\mathbin{?})}-operator non-deterministically chooses between two given arguments.
Non-determinism is not reflected in the type system, that is, a non-deter\-mi\-nistic choice has type \ensuremath{\Varid{a}\to \Varid{a}\to \Varid{a}}.
Such non-deterministic computations introduced by \ensuremath{(\mathbin{?})} describe two individual computation branches; one for the left argument and one for the right argument of \ensuremath{(\mathbin{?})}.

We could also define \ensuremath{\Varid{coin}} in Prolog-style by giving two rules for coin.

\begin{hscode}\SaveRestoreHook
\column{B}{@{}>{\hspre}l<{\hspost}@{}}%
\column{E}{@{}>{\hspre}l<{\hspost}@{}}%
\>[B]{}\Varid{coin}\mathbin{::}\Conid{Dist}\;\Conid{Bool}{}\<[E]%
\\
\>[B]{}\Varid{coin}\mathrel{=}\Conid{Dist}\;\Conid{True}\;\frac{\mathrm{1}}{\mathrm{2}}{}\<[E]%
\\
\>[B]{}\Varid{coin}\mathrel{=}\Conid{Dist}\;\Conid{False}\;\frac{\mathrm{1}}{\mathrm{2}}{}\<[E]%
\ColumnHook
\end{hscode}\resethooks

Both implementations can be used interchangeably since the \ensuremath{(\mathbin{?})}-operator is defined in the Prolog-style using two rules as well.

\begin{hscode}\SaveRestoreHook
\column{B}{@{}>{\hspre}l<{\hspost}@{}}%
\column{E}{@{}>{\hspre}l<{\hspost}@{}}%
\>[B]{}(\mathbin{?})\mathbin{::}\Varid{a}\to \Varid{a}\to \Varid{a}{}\<[E]%
\\
\>[B]{}\Varid{x}\mathbin{?}\Varid{y}\mathrel{=}\Varid{x}{}\<[E]%
\\
\>[B]{}\Varid{x}\mathbin{?}\Varid{y}\mathrel{=}\Varid{y}{}\<[E]%
\ColumnHook
\end{hscode}\resethooks

Printing an expression in the REPL\footnote{We visualize the interactions with the REPL using \ensuremath{\mkern-3mu\mathbin{>}} as prompt.} evaluates the non-deterministic computations, thus, yields one result for each branch as shown in the following examples.

\noindent
\begin{minipage}{0.48\textwidth}
\begin{hscode}\SaveRestoreHook
\column{B}{@{}>{\hspre}l<{\hspost}@{}}%
\column{E}{@{}>{\hspre}l<{\hspost}@{}}%
\>[B]{}\mkern-3mu\mathbin{>}\mathrm{1}\mathbin{?}\mathrm{2}{}\<[E]%
\\
\>[B]{}\mathrm{1}{}\<[E]%
\\
\>[B]{}\mathrm{2}{}\<[E]%
\ColumnHook
\end{hscode}\resethooks
\end{minipage}
\begin{minipage}{0.48\textwidth}
\begin{hscode}\SaveRestoreHook
\column{B}{@{}>{\hspre}l<{\hspost}@{}}%
\column{E}{@{}>{\hspre}l<{\hspost}@{}}%
\>[B]{}\mkern-3mu\mathbin{>}\Varid{coin}{}\<[E]%
\\
\>[B]{}\Conid{Dist}\;\Conid{True}\;\mathrm{0.5}{}\<[E]%
\\
\>[B]{}\Conid{Dist}\;\Conid{False}\;\mathrm{0.5}{}\<[E]%
\ColumnHook
\end{hscode}\resethooks
\end{minipage}

The REPL computes the values using a breadth-first-search strategy to visualize the results.
Due to the search strategy, we observe different outputs when changing the order of arguments to \ensuremath{(\mathbin{?})}.
However, because Curry's semantics is set-based \cite{christiansen2011adequate} the order of the results does not matter.

It is cumbersome to define distributions explicitly as in the case of \ensuremath{\Varid{coin}}.
Hence, we define helper functions for constructing distributions.\footnote{The definitions of predefined Curry functions like \ensuremath{\Varid{foldr}} are listed in \ref{appendix:functions}.}
Given a list of events and probabilities, \ensuremath{\Varid{enum}} creates a distribution by folding these pairs non-de\-ter\-min\-istically with a helper function \ensuremath{\Varid{member}}.\footnote{We shorten the implementation of \ensuremath{\Varid{enum}} for presentation purposes; actually, \ensuremath{\Varid{enum}} only allows valid distributions, e.g., that the given probabilities sum up to \ensuremath{\mathrm{1.0}}.}
\begin{hscode}\SaveRestoreHook
\column{B}{@{}>{\hspre}l<{\hspost}@{}}%
\column{3}{@{}>{\hspre}l<{\hspost}@{}}%
\column{E}{@{}>{\hspre}l<{\hspost}@{}}%
\>[3]{}\Varid{member}\mathbin{::}[\mskip1.5mu \Varid{a}\mskip1.5mu]\to \Varid{a}{}\<[E]%
\\
\>[3]{}\Varid{member}\;\Varid{xs}\mathrel{=}\Varid{foldr}\;(\mathbin{?})\;\Varid{failed}\;\Varid{xs}{}\<[E]%
\\[\blanklineskip]%
\>[3]{}\Varid{enum}\mathbin{::}[\mskip1.5mu \Varid{a}\mskip1.5mu]\to [\mskip1.5mu \Conid{Probability}\mskip1.5mu]\to \Conid{Dist}\;\Varid{a}{}\<[E]%
\\
\>[3]{}\Varid{enum}\;\Varid{xs}\;\Varid{ps}\mathrel{=}\Varid{member}\;(\Varid{zipWith}\;\Conid{Dist}\;\Varid{xs}\;\Varid{ps}){}\<[E]%
\ColumnHook
\end{hscode}\resethooks
In Curry the constant \ensuremath{\Varid{failed}} is a silent failure that behaves as neutral element with respect to \ensuremath{(\mathbin{?})}. %
That is, the expression \ensuremath{\Conid{True}\mathbin{?}\Varid{failed}} is equivalent to \ensuremath{\Conid{True}}. %
Hence, the function \ensuremath{\Varid{member}} takes a list and yields a non-deterministic choice of all elements of the list.

As a shortcut, we define a function that yields a \ensuremath{\Varid{uniform}} distribution given a list of events as well as a function \ensuremath{\Varid{certainly}}, which yields a distribution with a single event
of probability one.
\begin{hscode}\SaveRestoreHook
\column{B}{@{}>{\hspre}l<{\hspost}@{}}%
\column{3}{@{}>{\hspre}l<{\hspost}@{}}%
\column{E}{@{}>{\hspre}l<{\hspost}@{}}%
\>[3]{}\Varid{uniform}\mathbin{::}[\mskip1.5mu \Varid{a}\mskip1.5mu]\to \Conid{Dist}\;\Varid{a}{}\<[E]%
\\
\>[3]{}\Varid{uniform}\;\Varid{xs}\mathrel{=}\mathbf{let}\;\Varid{len}\mathrel{=}\Varid{length}\;\Varid{xs}\;\mathbf{in}\;\Varid{enum}\;\Varid{xs}\;(\Varid{repeat}\;\frac{\mathrm{1}}{\Varid{len}}){}\<[E]%
\\[\blanklineskip]%
\>[3]{}\Varid{certainly}\mathbin{::}\Varid{a}\to \Conid{Dist}\;\Varid{a}{}\<[E]%
\\
\>[3]{}\Varid{certainly}\;\Varid{x}\mathrel{=}\Conid{Dist}\;\Varid{x}\;\mathrm{1.0}{}\<[E]%
\ColumnHook
\end{hscode}\resethooks
The function \ensuremath{\Varid{repeat}} yields a list that contains the given value infinitely often.
Because of Curry's laziness, it is sufficient if one of the arguments of \ensuremath{\Varid{enum}} is a finite list because \ensuremath{\Varid{zipWith}} stops when one of its arguments is empty.
We can then refactor the definition of \ensuremath{\Varid{coin}} using \ensuremath{\Varid{uniform}} as follows.

\begin{hscode}\SaveRestoreHook
\column{B}{@{}>{\hspre}l<{\hspost}@{}}%
\column{E}{@{}>{\hspre}l<{\hspost}@{}}%
\>[B]{}\Varid{coin}\mathbin{::}\Conid{Dist}\;\Conid{Bool}{}\<[E]%
\\
\>[B]{}\Varid{coin}\mathrel{=}\Varid{uniform}\;[\mskip1.5mu \Conid{True},\Conid{False}\mskip1.5mu]{}\<[E]%
\ColumnHook
\end{hscode}\resethooks
In general, the library hides the constructor \ensuremath{\Conid{Dist}}, that is, the user has to define distributions by using the combinators provided by the library.

The library provides additional functions to combine and manipulate distributions.
In order to work with dependent distributions, the operator \ensuremath{(\mathbin{>\!\!\!>\!\!\!>\!=})} applies a function, which yields a distribution, to each event of a given distribution and multiplies the corresponding probabilities.\footnote{Due to the lack of overloading in Curry, operations on \ensuremath{\Conid{Float}} have a (floating) point suffix, e.g. \ensuremath{(\mathbin{*.})}, whereas operations on \ensuremath{\Conid{Int}} use the common operation names.}
\begin{hscode}\SaveRestoreHook
\column{B}{@{}>{\hspre}l<{\hspost}@{}}%
\column{3}{@{}>{\hspre}l<{\hspost}@{}}%
\column{15}{@{}>{\hspre}l<{\hspost}@{}}%
\column{20}{@{}>{\hspre}l<{\hspost}@{}}%
\column{E}{@{}>{\hspre}l<{\hspost}@{}}%
\>[3]{}(\mathbin{>\!\!\!>\!\!\!>\!=})\mathbin{::}\Conid{Dist}\;\Varid{a}\to (\Varid{a}\to \Conid{Dist}\;\Varid{b})\to \Conid{Dist}\;\Varid{b}{}\<[E]%
\\
\>[3]{}\Varid{d}\mathbin{>\!\!\!>\!\!\!>\!=}\Varid{f}\mathrel{=}{}\<[15]%
\>[15]{}\mathbf{let}\;{}\<[20]%
\>[20]{}\Conid{Dist}\;\Varid{x}\;\Varid{p}\mathrel{=}\Varid{d}{}\<[E]%
\\
\>[20]{}\Conid{Dist}\;\Varid{y}\;\Varid{q}\mathrel{=}\Varid{f}\;\Varid{x}{}\<[E]%
\\
\>[15]{}\mathbf{in}\;\Conid{Dist}\;\Varid{y}\;(\Varid{p}\mathbin{*.}\Varid{q}){}\<[E]%
\ColumnHook
\end{hscode}\resethooks
Intuitively, we have to apply the function \ensuremath{\Varid{f}} to each event of the distribution \ensuremath{\Varid{d}} and combine the resulting distributions into a single distribution.
In a Haskell implementation, we would use a list comprehension to define this function.
In the Curry implementation, we model distributions as non-deterministic computations, thus, the above rule describes the behavior of the function for an arbitrary pair of the first distribution and an arbitrary pair of the second distribution, that is, the result of \ensuremath{\Varid{f}}.

Using the operator \ensuremath{(\mathbin{>\!\!\!>\!\!\!>\!=})} we can, for example, define a distribution that models flipping two coins.
The events of this distribution are pairs whose first component is the result of the first coin flip and whose second component is the result of the second coin flip.
\begin{hscode}\SaveRestoreHook
\column{B}{@{}>{\hspre}l<{\hspost}@{}}%
\column{3}{@{}>{\hspre}l<{\hspost}@{}}%
\column{E}{@{}>{\hspre}l<{\hspost}@{}}%
\>[3]{}\Varid{independentCoins}\mathbin{::}\Conid{Dist}\;(\Conid{Bool},\Conid{Bool}){}\<[E]%
\\
\>[3]{}\Varid{independentCoins}\mathrel{=}\Varid{coin}\mathbin{>\!\!\!>\!\!\!>\!=}(\lambda \Varid{b1}\to \Varid{coin}\mathbin{>\!\!\!>\!\!\!>\!=}(\lambda \Varid{b2}\to \Varid{certainly}\;(\Varid{b1},\Varid{b2}))){}\<[E]%
\ColumnHook
\end{hscode}\resethooks

In contrast to the example \ensuremath{\Varid{independentCoins}} we can also use the operator \ensuremath{(\mathbin{>\!\!\!>\!\!\!>\!=})} to combine two distributions where we choose the second distribution on basis of the result of the first.
For example, we can define a distribution that models flipping two coins, but in this case we only flip a second coin if the first coin yields heads.
\begin{hscode}\SaveRestoreHook
\column{B}{@{}>{\hspre}l<{\hspost}@{}}%
\column{3}{@{}>{\hspre}l<{\hspost}@{}}%
\column{E}{@{}>{\hspre}l<{\hspost}@{}}%
\>[3]{}\Varid{dependentCoins}\mathbin{::}\Conid{Dist}\;\Conid{Bool}{}\<[E]%
\\
\>[3]{}\Varid{dependentCoins}\mathrel{=}\Varid{coin}\mathbin{>\!\!\!>\!\!\!>\!=}(\lambda \Varid{b}\to \mathbf{if}\;\Varid{b}\;\mathbf{then}\;\Varid{coin}\;\mathbf{else}\;\Varid{certainly}\;\Conid{False}){}\<[E]%
\ColumnHook
\end{hscode}\resethooks

The implementation of \ensuremath{(\mathbin{>\!\!\!>\!\!\!>\!=})} via \ensuremath{\mathbf{let}}-bindings seems a bit tedious, however, it is important that we define \ensuremath{(\mathbin{>\!\!\!>\!\!\!>\!=})} as it is.
The canonical implementation performs pattern matching on the first argument but uses a \ensuremath{\mathbf{let}}-binding for the result of \ensuremath{\Varid{f}}.
That is, it is strict in the first argument but non-strict in the application of \ensuremath{\Varid{f}}, the second argument.
For now it is sufficient to note --- and keep in mind --- that there is a difference between pattern matching and using \ensuremath{\mathbf{let}}-bindings.
In order to understand this difference, let us consider the following implementation of \ensuremath{\Varid{fromJustToList}} and an alternative implementation \ensuremath{\Varid{fromJustToListLet}}.\footnote{\ensuremath{(\mathbin{:})\mathbin{::}\Varid{a}\to [\mskip1.5mu \Varid{a}\mskip1.5mu]\to [\mskip1.5mu \Varid{a}\mskip1.5mu]} denotes the constructor for a non-empty list --- similar to the functor \emph{./2} in Prolog.}
\begin{hscode}\SaveRestoreHook
\column{B}{@{}>{\hspre}l<{\hspost}@{}}%
\column{3}{@{}>{\hspre}l<{\hspost}@{}}%
\column{E}{@{}>{\hspre}l<{\hspost}@{}}%
\>[3]{}\Varid{fromJustToList}\mathbin{::}\Conid{Maybe}\;\Varid{a}\to [\mskip1.5mu \Varid{a}\mskip1.5mu]{}\<[E]%
\\
\>[3]{}\Varid{fromJustToList}\;(\Conid{Just}\;\Varid{x})\mathrel{=}\Varid{x}\mathbin{:}[\mskip1.5mu \mskip1.5mu]{}\<[E]%
\ColumnHook
\end{hscode}\resethooks
\begin{hscode}\SaveRestoreHook
\column{B}{@{}>{\hspre}l<{\hspost}@{}}%
\column{3}{@{}>{\hspre}l<{\hspost}@{}}%
\column{E}{@{}>{\hspre}l<{\hspost}@{}}%
\>[3]{}\Varid{fromJustToListLet}\mathbin{::}\Conid{Maybe}\;\Varid{a}\to [\mskip1.5mu \Varid{a}\mskip1.5mu]{}\<[E]%
\\
\>[3]{}\Varid{fromJustToListLet}\;\Varid{mx}\mathrel{=}\mathbf{let}\;\Conid{Just}\;\Varid{x}\mathrel{=}\Varid{mx}\;\mathbf{in}\;\Varid{x}\mathbin{:}[\mskip1.5mu \mskip1.5mu]{}\<[E]%
\ColumnHook
\end{hscode}\resethooks

The second implementation, \ensuremath{\Varid{fromJustToListLet}}, is less strict, because it yields a list constructor, \ensuremath{(\mathbin{:})}, without evaluating its argument first.
That is, we can observe the difference when passing \ensuremath{\Varid{failed}} and checking if the resulting list is empty or not.

\begin{hscode}\SaveRestoreHook
\column{B}{@{}>{\hspre}l<{\hspost}@{}}%
\column{E}{@{}>{\hspre}l<{\hspost}@{}}%
\>[B]{}\mkern-3mu\mathbin{>}\Varid{null}\;(\Varid{fromJustToList}\;\Varid{failed}){}\<[E]%
\\
\>[B]{}\Varid{failed}{}\<[E]%
\\[\blanklineskip]%
\>[B]{}\mkern-3mu\mathbin{>}\Varid{null}\;(\Varid{fromJustToListLet}\;\Varid{failed}){}\<[E]%
\\
\>[B]{}\Conid{False}{}\<[E]%
\ColumnHook
\end{hscode}\resethooks

Due to the pattern matching in the definition of \ensuremath{\Varid{fromJustToList}} the argument \ensuremath{\Varid{failed}} needs to be evaluated and, thus, the function \ensuremath{\Varid{null}} propagates \ensuremath{\Varid{failed}} as result.
In contrast, the definition of \ensuremath{\Varid{fromJustToListLet}} postpones the evaluation of its argument to the right-hand side, i.e., the argument needs to be evaluated only if the computation demands the value \ensuremath{\Varid{x}} explicitly.
The function \ensuremath{\Varid{null}} does not demand the evaluation of \ensuremath{\Varid{x}}, because it only checks the surrounding list constructor.
\begin{hscode}\SaveRestoreHook
\column{B}{@{}>{\hspre}l<{\hspost}@{}}%
\column{3}{@{}>{\hspre}l<{\hspost}@{}}%
\column{18}{@{}>{\hspre}l<{\hspost}@{}}%
\column{E}{@{}>{\hspre}l<{\hspost}@{}}%
\>[3]{}\Varid{null}\mathbin{::}[\mskip1.5mu \Varid{a}\mskip1.5mu]\to \Conid{Bool}{}\<[E]%
\\
\>[3]{}\Varid{null}\;[\mskip1.5mu \mskip1.5mu]{}\<[18]%
\>[18]{}\mathrel{=}\Conid{True}{}\<[E]%
\\
\>[3]{}\Varid{null}\;(\Varid{x}\mathbin{:}\Varid{xs}){}\<[18]%
\>[18]{}\mathrel{=}\Conid{False}{}\<[E]%
\ColumnHook
\end{hscode}\resethooks

The same strictness property as for \ensuremath{\Varid{fromJustToList}} holds for a definition via explicit pattern matching using \ensuremath{\mathbf{case}\mathbin{...}\mathbf{of}}.
In particular, pattern matching of the left-hand side of a rule desugars to case expressions on the right-hand side.
\begin{hscode}\SaveRestoreHook
\column{B}{@{}>{\hspre}l<{\hspost}@{}}%
\column{3}{@{}>{\hspre}l<{\hspost}@{}}%
\column{33}{@{}>{\hspre}l<{\hspost}@{}}%
\column{E}{@{}>{\hspre}l<{\hspost}@{}}%
\>[3]{}\Varid{fromJustToListCase}\mathbin{::}\Conid{Maybe}\;\Varid{a}\to [\mskip1.5mu \Varid{a}\mskip1.5mu]{}\<[E]%
\\
\>[3]{}\Varid{fromJustToListCase}\;\Varid{mx}\mathrel{=}\mathbf{case}\;{}\<[33]%
\>[33]{}\Varid{mx}\;\mathbf{of}{}\<[E]%
\\
\>[33]{}\Conid{Just}\;\Varid{x}\to [\mskip1.5mu \Varid{x}\mskip1.5mu]{}\<[E]%
\ColumnHook
\end{hscode}\resethooks
\begin{hscode}\SaveRestoreHook
\column{B}{@{}>{\hspre}l<{\hspost}@{}}%
\column{E}{@{}>{\hspre}l<{\hspost}@{}}%
\>[B]{}\mkern-3mu\mathbin{>}\Varid{null}\;(\Varid{fromJustToListCase}\;\Varid{failed}){}\<[E]%
\\
\>[B]{}\Varid{failed}{}\<[E]%
\ColumnHook
\end{hscode}\resethooks

We discuss the implementation of \ensuremath{(\mathbin{>\!\!\!>\!\!\!>\!=})} in more detail later.
For now, it is sufficient to keep in mind that \ensuremath{(\mathbin{>\!\!\!>\!\!\!>\!=})} yields a \ensuremath{\Conid{Dist}}-constructor without evaluating any of its arguments.
In contrast, a definition using pattern matching or a case expression needs to evaluate its argument first, thus, is more strict.

For independent distributions we provide the function \ensuremath{\Varid{joinWith}} that combines two distributions with respect to a given function.
We implement \ensuremath{\Varid{joinWith}} by means of \ensuremath{(\mathbin{>\!\!\!>\!\!\!>\!=})}.
\begin{hscode}\SaveRestoreHook
\column{B}{@{}>{\hspre}l<{\hspost}@{}}%
\column{3}{@{}>{\hspre}l<{\hspost}@{}}%
\column{23}{@{}>{\hspre}l<{\hspost}@{}}%
\column{E}{@{}>{\hspre}l<{\hspost}@{}}%
\>[3]{}\Varid{joinWith}\mathbin{::}(\Varid{a}\to \Varid{b}\to \Varid{c})\to \Conid{Dist}\;\Varid{a}\to \Conid{Dist}\;\Varid{b}\to \Conid{Dist}\;\Varid{c}{}\<[E]%
\\
\>[3]{}\Varid{joinWith}\;\Varid{f}\;\Varid{d1}\;\Varid{d2}\mathrel{=}{}\<[23]%
\>[23]{}\Varid{d1}\mathbin{>\!\!\!>\!\!\!>\!=}(\lambda \Varid{x}\to \Varid{d2}\mathbin{>\!\!\!>\!\!\!>\!=}(\lambda \Varid{y}\to \Varid{certainly}\;(\Varid{f}\;\Varid{x}\;\Varid{y}))){}\<[E]%
\ColumnHook
\end{hscode}\resethooks
In a monadic setting this function is sometimes called \ensuremath{\Varid{liftM2}}.
Here, we use the same nomenclature as Erwig and Kollmansberger \cite{erwig2006functional}.

As an example we define a function that flips a coin $n$ times.
\begin{hscode}\SaveRestoreHook
\column{B}{@{}>{\hspre}l<{\hspost}@{}}%
\column{3}{@{}>{\hspre}l<{\hspost}@{}}%
\column{15}{@{}>{\hspre}l<{\hspost}@{}}%
\column{28}{@{}>{\hspre}l<{\hspost}@{}}%
\column{E}{@{}>{\hspre}l<{\hspost}@{}}%
\>[3]{}\Varid{flipCoin}\mathbin{::}\Conid{Int}\to \Conid{Dist}\;[\mskip1.5mu \Conid{Bool}\mskip1.5mu]{}\<[E]%
\\
\>[3]{}\Varid{flipCoin}\;\Varid{n}{}\<[15]%
\>[15]{}\mid \Varid{n}\equiv \mathrm{0}{}\<[28]%
\>[28]{}\mathrel{=}\Varid{certainly}\;[\mskip1.5mu \mskip1.5mu]{}\<[E]%
\\
\>[15]{}\mid \Varid{otherwise}{}\<[28]%
\>[28]{}\mathrel{=}\Varid{joinWith}\;(\mathbin{:})\;\Varid{coin}\;(\Varid{flipCoin}\;(\Varid{n}\mathbin{-}\mathrm{1})){}\<[E]%
\ColumnHook
\end{hscode}\resethooks
When we run the example of flipping two coins in the REPL of KiCS2, we get four events.

\begin{hscode}\SaveRestoreHook
\column{B}{@{}>{\hspre}l<{\hspost}@{}}%
\column{E}{@{}>{\hspre}l<{\hspost}@{}}%
\>[B]{}\mkern-3mu\mathbin{>}\Varid{flipCoin}\;\mathrm{2}{}\<[E]%
\\
\>[B]{}\Conid{Dist}\;[\mskip1.5mu \Conid{True},\Conid{True}\mskip1.5mu]\;\mathrm{0.25}{}\<[E]%
\\
\>[B]{}\Conid{Dist}\;[\mskip1.5mu \Conid{True},\Conid{False}\mskip1.5mu]\;\mathrm{0.25}{}\<[E]%
\\
\>[B]{}\Conid{Dist}\;[\mskip1.5mu \Conid{False},\Conid{True}\mskip1.5mu]\;\mathrm{0.25}{}\<[E]%
\\
\>[B]{}\Conid{Dist}\;[\mskip1.5mu \Conid{False},\Conid{False}\mskip1.5mu]\;\mathrm{0.25}{}\<[E]%
\ColumnHook
\end{hscode}\resethooks
In the example above, \ensuremath{\Varid{coin}} is non-deterministic, namely, \ensuremath{\Varid{coin}\mathrel{=}(\Conid{Dist}\;\Conid{True}\;\frac{\mathrm{1}}{\mathrm{2}})\mathbin{?}(\Conid{Dist}\;\Conid{False}\;\frac{\mathrm{1}}{\mathrm{2}})}.
Applying \ensuremath{\Varid{joinWith}} to \ensuremath{\Varid{coin}} and \ensuremath{\Varid{coin}} combines all possible results of two coin tosses.

\subsection{Querying Distributions}

With a handful of building blocks to define distributions available, we now want to query the distribution, that is, calculate the probability of certain events.
We provide an operator \ensuremath{(\mathbin{??})\mathbin{::}(\Varid{a}\to \Conid{Bool})\to \Conid{Dist}\;\Varid{a}\to \Conid{Probability}} --- which we will define shortly --- to extract the probability of an event.
The event is specified as a predicate, passed as first argument.
The operator filters events that satisfy the given predicate and computes the sum of the probabilities of the remaining elementary events.
We implement this kind of filter function on distributions in Curry.
\begin{hscode}\SaveRestoreHook
\column{B}{@{}>{\hspre}l<{\hspost}@{}}%
\column{3}{@{}>{\hspre}l<{\hspost}@{}}%
\column{24}{@{}>{\hspre}l<{\hspost}@{}}%
\column{E}{@{}>{\hspre}l<{\hspost}@{}}%
\>[3]{}\Varid{filterDist}\mathbin{::}(\Varid{a}\to \Conid{Bool})\to \Conid{Dist}\;\Varid{a}\to \Conid{Dist}\;\Varid{a}{}\<[E]%
\\
\>[3]{}\Varid{filterDist}\;\Varid{pred}\;\Varid{d}\mathrel{=}{}\<[24]%
\>[24]{}\mathbf{let}\;\Conid{Dist}\;\Varid{x}\;\Varid{p}\mathrel{=}\Varid{d}{}\<[E]%
\\
\>[24]{}\mathbf{in}\;\mathbf{if}\;(\Varid{pred}\;\Varid{x})\;\mathbf{then}\;(\Conid{Dist}\;\Varid{x}\;\Varid{p})\;\mathbf{else}\;\Varid{failed}{}\<[E]%
\ColumnHook
\end{hscode}\resethooks
The implementation of \ensuremath{\Varid{filterDist}} is a partial identity on the event-probability pairs.
Every event that satisfies the predicate is part of the resulting distribution.
The function fails for event-probability pairs that do not satisfy the predicate.

Querying a distribution, i.e., summing up all probabilities that satisfy a predicate, is a more advanced task in the functional logic approach.
Remember that we represent a distribution by chaining all event-probability pairs with \ensuremath{(\mathbin{?})}, thus, constructing non-deterministic computations.
These non-deterministic computations introduce individual branches of computations that cannot interact with each other. %
In order to compute the total probability of a distribution, we have to merge these distinct branches. %
Such a merge is possible by the encapsulation of non-deterministic computations. %
Similar to the \emph{findall} construct of the logic language Prolog, in Curry we encapsulate a non-deterministic computation by using a primitive called \ensuremath{\Varid{allValues}}\footnote{We use an abstract view of the result of an encapsulation to emphasize that the order of encapsulated results does not matter. In practice, we can, for example, use the function \ensuremath{\Varid{allValues}\mathbin{::}\Varid{a}\to [\mskip1.5mu \Varid{a}\mskip1.5mu]} defined in the library \ensuremath{\Conid{Findall}}.}. %
The function \ensuremath{\Varid{allValues}\mathbin{::}\Varid{a}\to \{\Varid{a}\}} operates on a polymorphic --- and potentially non-deterministic --- value and yields a multiset of all non-deterministic values. %
In order to work with encapsulated values, Curry provides the function \ensuremath{\Varid{foldValues}\mathbin{::}(\Varid{a}\to \Varid{a}\to \Varid{a})\to \Varid{a}\to \{\Varid{a}\}\to \Varid{a}} to fold the resulting multiset.

We do not discuss the implementation details behind \ensuremath{\Varid{allValues}} here.
It is sufficient to know that, as a library developer, we can employ this function to encapsulate
non-deterministic values and use these values in further computations.
However, due to non-transparent behavior in combination with sharing as discussed by Braßel et al. \cite{brassel2004encapsulating}, a user of the library should not use \ensuremath{\Varid{allValues}} at all. In a nutshell, inner-most and outer-most evaluation strategies may cause different results when combining sharing and encapsulation.

With this encapsulation mechanism at hand, we can define the query operator \ensuremath{(\mathbin{??})} as follows. %
\begin{hscode}\SaveRestoreHook
\column{B}{@{}>{\hspre}l<{\hspost}@{}}%
\column{3}{@{}>{\hspre}l<{\hspost}@{}}%
\column{E}{@{}>{\hspre}l<{\hspost}@{}}%
\>[3]{}\Varid{prob}\mathbin{::}\Conid{Dist}\;\Varid{a}\to \Conid{Probability}{}\<[E]%
\\
\>[3]{}\Varid{prob}\;(\Conid{Dist}\;\Varid{x}\;\Varid{p})\mathrel{=}\Varid{p}{}\<[E]%
\\[\blanklineskip]%
\>[3]{}(\mathbin{??})\mathbin{::}(\Varid{a}\to \Conid{Bool})\to \Conid{Dist}\;\Varid{a}\to \Conid{Probability}{}\<[E]%
\\
\>[3]{}(\mathbin{??})\;\Varid{pred}\;\Varid{d}\mathrel{=}\Varid{foldValues}\;(\mathbin{+.})\;\mathrm{0.0}\;(\Varid{allValues}\;(\Varid{prob}\;(\Varid{filterDist}\;\Varid{pred}\;\Varid{d}))){}\<[E]%
\ColumnHook
\end{hscode}\resethooks
First we filter the elementary events by some predicate and project to the probabilities only.
Afterwards we encapsulate the remaining probabilities and sum them up.
As an example for the use of \ensuremath{(\mathbin{??})}, we may flip four coins and calculate the probability of at least two heads --- that is, the list contains at least two \ensuremath{\Conid{True}} values.

\begin{hscode}\SaveRestoreHook
\column{B}{@{}>{\hspre}l<{\hspost}@{}}%
\column{E}{@{}>{\hspre}l<{\hspost}@{}}%
\>[B]{}\mkern-3mu\mathbin{>}(\lambda \Varid{coins}\to \Varid{length}\;(\Varid{filter}\;\Varid{id}\;\Varid{coins})\geq \mathrm{2})\mathbin{??}(\Varid{flipCoin}\;\mathrm{4}){}\<[E]%
\\
\>[B]{}\mathrm{0.6875}{}\<[E]%
\ColumnHook
\end{hscode}\resethooks

In order to check the result, we calculate the probability by hand.
Since there are more events that satisfy the predicate than events that do not, we sum up the probabilities of the events that do not satisfy the predicate and calculate the complementary probability.
There is one event where all coins show tails and four events where one of the coins shows heads and all other show tails.
\begin{align*}
  & 1 - ( P(Tails) \cdot P(Tails) \cdot P(Tails) \cdot P(Tails) + 4 \cdot P(Heads) \cdot P(Tails) \cdot P(Tails) \cdot P(Tails))\\
=~ & 1 - (0.5 \cdot 0.5 \cdot 0.5 \cdot 0.5 + 4 \cdot 0.5 \cdot 0.5 \cdot 0.5 \cdot 0.5)\\
=~ & 1 - (0.0625 + 0.25) = 1 - 0.3125 = 0.6875
\end{align*}

\section{Library Pragmatics}
\label{sec:details}

Up to now, we have discussed a simple library for probabilistic programming that uses non-determinism to represent distributions.
In this section we will see that we can highly benefit from Curry-like non-determinism with respect to performance when we compare PFLP's implementation with a list-based implementation.
More precisely, when we query a distribution with a predicate that does not evaluate its argument completely, we can possibly prune large parts of the search space.
Before we discuss the details of the combination of non-strictness and non-determinism, we discuss aspects of sharing non-deterministic choices.
Finally, we discuss details about the implementation of \ensuremath{(\mathbin{>\!\!\!>\!\!\!>\!=})} and why PFLP does not allow non-deterministic events within distributions.

\subsection{Call-time Choice vs. Run-time Choice}

By default Curry uses call-time choice, that is, variables denote single deterministic choices.
When we bind a variable to a non-deterministic computation, one value is chosen and all occurrences of the variable denote the same deterministic choice.
Often call-time choice is what you are looking for.
For example, this slightly modified definition of \ensuremath{\Varid{filterDist}} makes use of call-time choice.
\begin{hscode}\SaveRestoreHook
\column{B}{@{}>{\hspre}l<{\hspost}@{}}%
\column{22}{@{}>{\hspre}l<{\hspost}@{}}%
\column{E}{@{}>{\hspre}l<{\hspost}@{}}%
\>[B]{}\Varid{filterDist}\mathbin{::}(\Varid{a}\to \Conid{Bool})\to \Conid{Dist}\;\Varid{a}\to \Conid{Dist}\;\Varid{a}{}\<[E]%
\\
\>[B]{}\Varid{filterDist}\;\Varid{pred}\;\Varid{d}\mathrel{=}{}\<[22]%
\>[22]{}\mathbf{let}\;\Conid{Dist}\;\Varid{x}\;\Varid{p}\mathrel{=}\Varid{d}{}\<[E]%
\\
\>[22]{}\mathbf{in}\;\mathbf{if}\;(\Varid{pred}\;\Varid{x})\;\mathbf{then}\;\Varid{d}\;\mathbf{else}\;\Varid{failed}{}\<[E]%
\ColumnHook
\end{hscode}\resethooks
Due to pattern matching via \ensuremath{\mathbf{let}}-binding the variable \ensuremath{\Varid{d}} on the right-hand side corresponds to a single deterministic choice for the input distribution, namely, the one that satisfies the predicate and not the non-deterministic computation that was initially passed as second argument to \ensuremath{\Varid{filterDist}}.

Almost as often run-time choice is what you are looking for and call-time choice gets in your way; probabilistic programming is no exception.
For example, let us reconsider flipping a coin $n$ times.
We parametrize the function \ensuremath{\Varid{flipCoin}} over the given distribution and define the following generalized function.
\begin{hscode}\SaveRestoreHook
\column{B}{@{}>{\hspre}l<{\hspost}@{}}%
\column{20}{@{}>{\hspre}l<{\hspost}@{}}%
\column{33}{@{}>{\hspre}l<{\hspost}@{}}%
\column{E}{@{}>{\hspre}l<{\hspost}@{}}%
\>[B]{}\Varid{replicateDist}\mathbin{::}\Conid{Int}\to \Conid{Dist}\;\Varid{a}\to \Conid{Dist}\;[\mskip1.5mu \Varid{a}\mskip1.5mu]{}\<[E]%
\\
\>[B]{}\Varid{replicateDist}\;\Varid{n}\;\Varid{d}{}\<[20]%
\>[20]{}\mid \Varid{n}\equiv \mathrm{0}{}\<[33]%
\>[33]{}\mathrel{=}\Varid{certainly}\;[\mskip1.5mu \mskip1.5mu]{}\<[E]%
\\
\>[20]{}\mid \Varid{otherwise}{}\<[33]%
\>[33]{}\mathrel{=}\Varid{joinWith}\;(\mathbin{:})\;\Varid{d}\;(\Varid{replicateDist}\;(\Varid{n}\mathbin{-}\mathrm{1})\;\Varid{d}){}\<[E]%
\ColumnHook
\end{hscode}\resethooks

Here, we again use guard syntax in order to distinguish several cases depending on the Boolean expression.
When we use this function to flip a coin twice, the result is not what we intended. %
\begin{hscode}\SaveRestoreHook
\column{B}{@{}>{\hspre}l<{\hspost}@{}}%
\column{E}{@{}>{\hspre}l<{\hspost}@{}}%
\>[B]{}\mkern-3mu\mathbin{>}\Varid{replicateDist}\;\mathrm{2}\;\Varid{coin}{}\<[E]%
\\
\>[B]{}\Conid{Dist}\;[\mskip1.5mu \Conid{True},\Conid{True}\mskip1.5mu]\;\mathrm{0.25}{}\<[E]%
\\
\>[B]{}\Conid{Dist}\;[\mskip1.5mu \Conid{False},\Conid{False}\mskip1.5mu]\;\mathrm{0.25}{}\<[E]%
\ColumnHook
\end{hscode}\resethooks
Because \ensuremath{\Varid{replicateDist}} shares the variable \ensuremath{\Varid{d}}, we only perform a choice once and replicate deterministic choices.
In contrast, top-level nullary functions like \ensuremath{\Varid{coin}} are evaluated every time, thus, exhibit run-time choice, which is the reason why the previously shown \ensuremath{\Varid{flipCoin}} behaves properly.

In order to implement \ensuremath{\Varid{replicateDist}} correctly, we have to enforce run-time choice.
We introduce the following type synonym and function to model and work with values with run-time choice behavior.
\begin{hscode}\SaveRestoreHook
\column{B}{@{}>{\hspre}l<{\hspost}@{}}%
\column{3}{@{}>{\hspre}l<{\hspost}@{}}%
\column{E}{@{}>{\hspre}l<{\hspost}@{}}%
\>[3]{}\mathbf{type}\;\Conid{RT}\;\Varid{a}\mathrel{=}()\to \Varid{a}{}\<[E]%
\\[\blanklineskip]%
\>[3]{}\Varid{pick}\mathbin{::}\Conid{RT}\;\Varid{a}\to \Varid{a}{}\<[E]%
\\
\>[3]{}\Varid{pick}\;\Varid{rt}\mathrel{=}\Varid{rt}\;(){}\<[E]%
\ColumnHook
\end{hscode}\resethooks
We can now use the type \ensuremath{\Conid{RT}} to hide the non-determinism on the right-hand side of a function arrow.
This way, \ensuremath{\Varid{pick}} explicitly triggers the evaluation of \ensuremath{\Varid{rt}}, performing a new choice for every element of the result list.
\begin{hscode}\SaveRestoreHook
\column{B}{@{}>{\hspre}l<{\hspost}@{}}%
\column{3}{@{}>{\hspre}l<{\hspost}@{}}%
\column{23}{@{}>{\hspre}l<{\hspost}@{}}%
\column{36}{@{}>{\hspre}l<{\hspost}@{}}%
\column{E}{@{}>{\hspre}l<{\hspost}@{}}%
\>[3]{}\Varid{replicateDist}\mathbin{::}\Conid{Int}\to \Conid{RT}\;(\Conid{Dist}\;\Varid{a})\to \Conid{Dist}\;[\mskip1.5mu \Varid{a}\mskip1.5mu]{}\<[E]%
\\
\>[3]{}\Varid{replicateDist}\;\Varid{n}\;\Varid{rt}{}\<[23]%
\>[23]{}\mid \Varid{n}\equiv \mathrm{0}{}\<[36]%
\>[36]{}\mathrel{=}\Varid{certainly}\;[\mskip1.5mu \mskip1.5mu]{}\<[E]%
\\
\>[23]{}\mid \Varid{otherwise}{}\<[36]%
\>[36]{}\mathrel{=}\Varid{joinWith}\;(\mathbin{:})\;(\Varid{pick}\;\Varid{rt})\;(\Varid{replicateDist}\;(\Varid{n}\mathbin{-}\mathrm{1})\;\Varid{rt}){}\<[E]%
\ColumnHook
\end{hscode}\resethooks

In order to use \ensuremath{\Varid{replicateDist}} with \ensuremath{\Varid{coin}}, we have to construct a value of type \ensuremath{\Conid{RT}\;(\Conid{Dist}\;\Conid{Bool})}.
However, we cannot provide a function to construct a value of type \ensuremath{\Conid{RT}} that behaves as intended.
Such a function would share a deterministic choice and non-de\-ter\-min\-istically yield two functions, instead of one function that yields a non-deterministic computation.
The only way to construct a value of type \ensuremath{\Conid{RT}} is to explicitly use a lambda abstraction.

\begin{hscode}\SaveRestoreHook
\column{B}{@{}>{\hspre}l<{\hspost}@{}}%
\column{E}{@{}>{\hspre}l<{\hspost}@{}}%
\>[B]{}\mkern-3mu\mathbin{>}\Varid{replicateDist}\;\mathrm{2}\;(\lambda ()\to \Varid{coin}){}\<[E]%
\\
\>[B]{}\Conid{Dist}\;[\mskip1.5mu \Conid{True},\Conid{True}\mskip1.5mu]\;\mathrm{0.25}{}\<[E]%
\\
\>[B]{}\Conid{Dist}\;[\mskip1.5mu \Conid{True},\Conid{False}\mskip1.5mu]\;\mathrm{0.25}{}\<[E]%
\\
\>[B]{}\Conid{Dist}\;[\mskip1.5mu \Conid{False},\Conid{True}\mskip1.5mu]\;\mathrm{0.25}{}\<[E]%
\\
\>[B]{}\Conid{Dist}\;[\mskip1.5mu \Conid{False},\Conid{False}\mskip1.5mu]\;\mathrm{0.25}{}\<[E]%
\ColumnHook
\end{hscode}\resethooks

Instead of relying on call-time choice as default behavior, we could model \ensuremath{\Conid{Dist}} as a function and make run-time choice the default in PFLP.
In this case, to get call-time choice we would have to use a special construct provided by the library --- as it is the case in many probabilistic programming libraries, e.g., \emph{mem} in WebPPL \cite{goodman2014design}.

On the other hand, ProbLog uses a similar concept to call-time choice, namely, stochastic memoization, which reuses already computed results.
That is, predicates that are associated with probabilities become part of the memoized result.
If a fair coin flip, for example, already resulted in \ensuremath{\Conid{True}}, then the probability of all further coin flips that also result in \ensuremath{\Conid{True}} have probability \ensuremath{\mathrm{1}}.
Due to stochastic memoization the coin is not flipped a second time, but is identified as the same coin as before.
Thus, stochastic memoization as used in ProbLog is similar to the extension of tabling in Prolog systems, but adapted to the setting of probabilistic programming that extends predicates with probabilities.
Similar to our usage of \ensuremath{\Conid{RT}} to mimic run-time choice in Curry, we can use a so-called trial identifier, which is basically an additional argument, to circumvent memoization for a predicate like coin in ProbLog.
The difference to \ensuremath{\Conid{RT}} is that the trial identifier needs to be different for each call to the predicate in order to force re-evaluation.

In the end, we have decided to go with the current modeling based on call-time choice, because the alternative would work against the spirit of the Curry programming language.
There is a long history of discussions about the pros and cons of call-time choice and run-time choice.
It is common knowledge in probabilistic programming~\cite{deraedt2013probabilistic} that memoization --- that is, call-time choice --- has to be avoided in order to model stochastic automata or probabilistic grammars.
Similarly,~Antoy \cite{antoy2005evaluation} observes that you need run-time choice to elegantly model regular expressions in the context of functional logic programming languages.
Then again, probabilistic languages need a concept like memoization in order to use a single value drawn from a distribution multiple times.

\subsection{Combination of Non-strictness and Non-determinism}
\label{ssec:nonstrict}

This subsection illustrates the benefits from the combination of non-strictness and non-deter\-mi\-nism with respect to performance.
More precisely, in a setting that uses Curry-like non-determinism, non-strictness can prevent non-determinism from being \enquote{spawned}.
Let us consider calculating the probability for throwing only sixes when throwing \ensuremath{\Varid{n}} dice.
First we define a uniform die as follows.
\begin{hscode}\SaveRestoreHook
\column{B}{@{}>{\hspre}l<{\hspost}@{}}%
\column{3}{@{}>{\hspre}l<{\hspost}@{}}%
\column{E}{@{}>{\hspre}l<{\hspost}@{}}%
\>[3]{}\mathbf{data}\;\Conid{Side}\mathrel{=}\Conid{One}\mid \Conid{Two}\mid \Conid{Three}\mid \Conid{Four}\mid \Conid{Five}\mid \Conid{Six}{}\<[E]%
\\[\blanklineskip]%
\>[3]{}\Varid{die}\mathbin{::}\Conid{Dist}\;\Conid{Side}{}\<[E]%
\\
\>[3]{}\Varid{die}\mathrel{=}\Varid{uniform}\;[\mskip1.5mu \Conid{One},\Conid{Two},\Conid{Three},\Conid{Four},\Conid{Five},\Conid{Six}\mskip1.5mu]{}\<[E]%
\ColumnHook
\end{hscode}\resethooks

We define the following query by means of the combinators introduced so far.
The function \ensuremath{\Varid{all}} simply checks that all elements of a list satisfy a given predicate; it is defined by means of the Boolean conjunction \ensuremath{(\mathrel{\wedge})}.
\begin{hscode}\SaveRestoreHook
\column{B}{@{}>{\hspre}l<{\hspost}@{}}%
\column{3}{@{}>{\hspre}l<{\hspost}@{}}%
\column{E}{@{}>{\hspre}l<{\hspost}@{}}%
\>[3]{}\Varid{allSix}\mathbin{::}\Conid{Int}\to \Conid{Probability}{}\<[E]%
\\
\>[3]{}\Varid{allSix}\;\Varid{n}\mathrel{=}(\Varid{all}\;(\equiv \Conid{Six}))\mathbin{??}(\Varid{replicateDist}\;\Varid{n}\;(\lambda ()\to \Varid{die})){}\<[E]%
\ColumnHook
\end{hscode}\resethooks

\autoref{tab:allSix} compares running times\footnote{All benchmarks were executed on a Linux machine with an Intel Core i7-6500U (2.50 GHz) and 8 GiB RAM running Fedora 25. We used the Glasgow Haskell Compiler (version \text{\tt 8\char46{}0\char46{}2}, option \text{\tt \char45{}O2}) and set the search strategy in KiCS2 to depth-first.} of this query for different numbers of dice.
The row labeled \enquote{Curry ND} lists the running times for an implementation that uses the operator \ensuremath{(\mathbin{>\!\!\!>\!\!\!>\!=})}.
The row \enquote{Curry List} shows the numbers for a list-based implementation in Curry, which is a literal translation of the library by Erwig and Kollmansberger.
The row labeled \enquote{Curry ND!} uses an operator \ensuremath{(\mathbin{>\!\!\!>\!\!\!>\!=!})} instead, which we will discuss shortly.
Finally, we compare our implementation to the original list-based implementation, which the row labeled \enquote{Haskell List} refers to.
The table states the running times in milliseconds of a compiled executable for each benchmark as a mean of three runs. Cells marked with \enquote{--} take more than one minute.

\begin{table*}[h]\centering
\begin{tabular}{@{}lrrrrrrrrr}
\toprule
\# of dice & 5 & 6 & 7 & 8 & 9 & 10 & 100 & 200 & 300\\
\midrule
Curry ND       & $<$\num{1} & $<$\num{1} & $<$\num{1} & $<$\num{1} & $<$\num{1} & $<$\num{1} & \num{48} & \num{231} & \num{547} \\
Curry List     & \num{2} & \num{13} & \num{72} & \num{419} & \num{2554} & \num{15394} & -- & -- & -- \\
Curry ND!      & \num{52} & \num{409} & \num{2568} & \num{16382} & -- & -- & -- & -- & -- \\
Haskell List   & \num{1} & \num{5} & \num{30} & \num{210} & \num{1415} & \num{6538} & -- & -- & -- \\
\bottomrule
\end{tabular}
\caption{Overview of running times for the query \ensuremath{\Varid{allSix}\;\Varid{n}}}
\label{tab:allSix}
\end{table*}

Obviously, the example above is a little contrived.
While the query is exponential in both list versions, it is linear in the non-deterministic setting\footnote{Non-determinism causes significant overhead for KiCS2, thus, \enquote{Curry ND} does not show linear development, but we measured a linear running time using PAKCS~\cite{hanus2017pakcs}.}.
To illustrate the behavior of the example above, we consider the following application for an arbitrary distribution \ensuremath{\Varid{dist}} of type \ensuremath{\Conid{Dist}\;[\mskip1.5mu \Conid{Side}\mskip1.5mu]}.
\begin{hscode}\SaveRestoreHook
\column{B}{@{}>{\hspre}l<{\hspost}@{}}%
\column{E}{@{}>{\hspre}l<{\hspost}@{}}%
\>[B]{}\Varid{filterDist}\;(\Varid{all}\;(\equiv \Conid{Six}))\;(\Varid{joinWith}\;(\mathbin{:})\;(\Conid{Dist}\;\Conid{One}\;\frac{\mathrm{1}}{\mathrm{6}})\;\Varid{dist}){}\<[E]%
\ColumnHook
\end{hscode}\resethooks

This application yields an empty distribution without evaluating the distribution \ensuremath{\Varid{dist}}.
The interesting point here is that \ensuremath{\Varid{joinWith}} yields a \ensuremath{\Conid{Dist}}-constructor without inspecting its arguments.
When we demand the event of the resulting \ensuremath{\Conid{Dist}}, \ensuremath{\Varid{joinWith}} has to evaluate only its first argument to see that the predicate \ensuremath{\Varid{all}\;(\equiv \Conid{Six})} yields \ensuremath{\Conid{False}}.
The evaluation of the expression fails without inspecting the second argument of \ensuremath{\Varid{joinWith}}.
\autoref{fig:eval} illustrates the evaluation in more detail.

\begin{figure}[t]
\centering
\begin{hscode}\SaveRestoreHook
\column{B}{@{}>{\hspre}l<{\hspost}@{}}%
\column{6}{@{}>{\hspre}l<{\hspost}@{}}%
\column{13}{@{}>{\hspre}l<{\hspost}@{}}%
\column{16}{@{}>{\hspre}l<{\hspost}@{}}%
\column{20}{@{}>{\hspre}l<{\hspost}@{}}%
\column{21}{@{}>{\hspre}l<{\hspost}@{}}%
\column{E}{@{}>{\hspre}l<{\hspost}@{}}%
\>[B]{}\Varid{filterDist}\;(\Varid{all}\;(\equiv \Conid{Six}))\;(\Varid{joinWith}\;(\mathbin{:})\;(\Conid{Dist}\;\Conid{One}\;\frac{\mathrm{1}}{\mathrm{6}})\;\Varid{dist}){}\<[E]%
\\
\>[B]{}\equiv \mbox{\commentbegin  Definition of \ensuremath{\Varid{joinWith}}  \commentend}{}\<[E]%
\\
\>[B]{}\Varid{filterDist}\;{}\<[13]%
\>[13]{}(\Varid{all}\;(\equiv \Conid{Six}))\;{}\<[E]%
\\
\>[13]{}(\Conid{Dist}\;\Conid{One}\;\frac{\mathrm{1}}{\mathrm{6}}\mathbin{>\!\!\!>\!\!\!>\!=}(\lambda \Varid{x}\to \Varid{dist}\mathbin{>\!\!\!>\!\!\!>\!=}(\lambda \Varid{xs}\to \Varid{certainly}\;(\Varid{x}\mathbin{:}\Varid{xs})))){}\<[E]%
\\
\>[B]{}\equiv \mbox{\commentbegin  Definition of \ensuremath{(\mathbin{>\!\!\!>\!\!\!>\!=})} (twice)  \commentend}{}\<[E]%
\\
\>[B]{}\Varid{filterDist}\;{}\<[13]%
\>[13]{}(\Varid{all}\;(\equiv \Conid{Six}))\;{}\<[E]%
\\
\>[13]{}({}\<[16]%
\>[16]{}\mathbf{let}\;{}\<[21]%
\>[21]{}\Conid{Dist}\;\Varid{x}\;\Varid{p}\mathrel{=}\Conid{Dist}\;\Conid{One}\;\frac{\mathrm{1}}{\mathrm{6}}{}\<[E]%
\\
\>[21]{}\Conid{Dist}\;\Varid{xs}\;\Varid{q}\mathrel{=}\Varid{dist}{}\<[E]%
\\
\>[21]{}\Conid{Dist}\;\Varid{ys}\;\Varid{r}\mathrel{=}\Varid{certainly}\;(\Varid{x}\mathbin{:}\Varid{xs}){}\<[E]%
\\
\>[16]{}\mathbf{in}\;{}\<[20]%
\>[20]{}\Conid{Dist}\;\Varid{ys}\;(\Varid{p}\mathbin{*.}(\Varid{q}\mathbin{*.}\Varid{r})){}\<[E]%
\\
\>[B]{}\equiv \mbox{\commentbegin  Definition of \ensuremath{\Varid{filterDist}}  \commentend}{}\<[E]%
\\
\>[B]{}\mathbf{let}\;{}\<[6]%
\>[6]{}\Conid{Dist}\;\Varid{x}\;\Varid{p}\mathrel{=}\Conid{Dist}\;\Conid{One}\;\frac{\mathrm{1}}{\mathrm{6}}{}\<[E]%
\\
\>[6]{}\Conid{Dist}\;\Varid{xs}\;\Varid{q}\mathrel{=}\Varid{dist}{}\<[E]%
\\
\>[6]{}\Conid{Dist}\;\Varid{ys}\;\Varid{r}\mathrel{=}\Varid{certainly}\;(\Varid{x}\mathbin{:}\Varid{xs}){}\<[E]%
\\
\>[B]{}\mathbf{in}\;\mathbf{if}\;\Varid{all}\;(\equiv \Conid{Six})\;\Varid{ys}\;\mathbf{then}\;\Conid{Dist}\;\Varid{ys}\;(\Varid{p}\mathbin{*.}(\Varid{q}\mathbin{*.}\Varid{r}))\;\mathbf{else}\;\Varid{failed}{}\<[E]%
\\
\>[B]{}\equiv \mbox{\commentbegin  Definition of \ensuremath{\Varid{certainly}}  \commentend}{}\<[E]%
\\
\>[B]{}\mathbf{let}\;{}\<[6]%
\>[6]{}\Conid{Dist}\;\Varid{x}\;\Varid{p}\mathrel{=}\Conid{Dist}\;\Conid{One}\;\frac{\mathrm{1}}{\mathrm{6}}{}\<[E]%
\\
\>[6]{}\Conid{Dist}\;\Varid{xs}\;\Varid{q}\mathrel{=}\Varid{dist}{}\<[E]%
\\
\>[B]{}\mathbf{in}\;\mathbf{if}\;\Varid{all}\;(\equiv \Conid{Six})\;(\Varid{x}\mathbin{:}\Varid{xs})\;\mathbf{then}\;\Conid{Dist}\;(\Varid{x}\mathbin{:}\Varid{xs})\;(\Varid{p}\mathbin{*.}(\Varid{q}\mathbin{*.}\mathrm{1.0}))\;\mathbf{else}\;\Varid{failed}{}\<[E]%
\\
\>[B]{}\equiv \mbox{\commentbegin  Definition of \ensuremath{\Varid{all}}  \commentend}{}\<[E]%
\\
\>[B]{}\mathbf{let}\;{}\<[6]%
\>[6]{}\Conid{Dist}\;\Varid{x}\;\Varid{p}\mathrel{=}\Conid{Dist}\;\Conid{One}\;\frac{\mathrm{1}}{\mathrm{6}}{}\<[E]%
\\
\>[6]{}\Conid{Dist}\;\Varid{xs}\;\Varid{q}\mathrel{=}\Varid{dist}{}\<[E]%
\\
\>[B]{}\mathbf{in}\;\mathbf{if}\;\Varid{x}\equiv \Conid{Six}\mathrel{\wedge}\Varid{all}\;(\equiv \Conid{Six})\;\Varid{xs}\;\mathbf{then}\;\Conid{Dist}\;(\Varid{x}\mathbin{:}\Varid{xs})\;(\Varid{p}\mathbin{*.}(\Varid{q}\mathbin{*.}\mathrm{1.0}))\;\mathbf{else}\;\Varid{failed}{}\<[E]%
\\
\>[B]{}\equiv \mbox{\commentbegin  Definition of \ensuremath{(\equiv )} and \ensuremath{(\mathrel{\wedge})}  \commentend}{}\<[E]%
\\
\>[B]{}\mathbf{let}\;{}\<[6]%
\>[6]{}\Conid{Dist}\;\Varid{x}\;\Varid{p}\mathrel{=}\Conid{Dist}\;\Conid{One}\;\frac{\mathrm{1}}{\mathrm{6}}{}\<[E]%
\\
\>[6]{}\Conid{Dist}\;\Varid{xs}\;\Varid{q}\mathrel{=}\Varid{d}{}\<[E]%
\\
\>[B]{}\mathbf{in}\;\mathbf{if}\;\Conid{False}\;\mathbf{then}\;\Conid{Dist}\;(\Varid{x}\mathbin{:}\Varid{xs})\;(\Varid{p}\mathbin{*.}(\Varid{q}\mathbin{*.}\mathrm{1.0}))\;\mathbf{else}\;\Varid{failed}{}\<[E]%
\\
\>[B]{}\equiv \mbox{\commentbegin  Definition of \ensuremath{\mathbf{if}\mathbin{-}\mathbf{then}\mathbin{-}\mathbf{else}}  \commentend}{}\<[E]%
\\
\>[B]{}\Varid{failed}{}\<[E]%
\ColumnHook
\end{hscode}\resethooks
\caption{Simplified evaluation illustrating non-strict non-determinism}
\label{fig:eval}
\end{figure}

In case of the example \ensuremath{\Varid{allSix}}, all non-deterministic branches that contain a value different from \ensuremath{\Conid{Six}} fail fast due to the non-strictness.
Thus, the number of evaluation steps is linear in the number of rolled dice.

We can only benefit from the combination of non-strictness and non-deter\-mi\-nism if we define \ensuremath{(\mathbin{>\!\!\!>\!\!\!>\!=})} with care.
Let us take a look at a strict variant of \ensuremath{(\mathbin{>\!\!\!>\!\!\!>\!=})} and discuss its consequences.
\begin{hscode}\SaveRestoreHook
\column{B}{@{}>{\hspre}l<{\hspost}@{}}%
\column{3}{@{}>{\hspre}l<{\hspost}@{}}%
\column{30}{@{}>{\hspre}l<{\hspost}@{}}%
\column{E}{@{}>{\hspre}l<{\hspost}@{}}%
\>[3]{}(\mathbin{>\!\!\!>\!\!\!>\!=!})\mathbin{::}\Conid{Dist}\;\Varid{a}\to (\Varid{a}\to \Conid{Dist}\;\Varid{b})\to \Conid{Dist}\;\Varid{b}{}\<[E]%
\\
\>[3]{}(\Conid{Dist}\;\Varid{x}\;\Varid{p})\mathbin{>\!\!\!>\!\!\!>\!=!}\Varid{f}\mathrel{=}\mathbf{case}\;{}\<[30]%
\>[30]{}\Varid{f}\;\Varid{x}\;\mathbf{of}{}\<[E]%
\\
\>[30]{}\Conid{Dist}\;\Varid{y}\;\Varid{q}\to \Conid{Dist}\;\Varid{y}\;(\Varid{p}\mathbin{*.}\Varid{q}){}\<[E]%
\ColumnHook
\end{hscode}\resethooks
This implementation is strict in its first argument as well as in the result of the function application.
When we use \ensuremath{(\mathbin{>\!\!\!>\!\!\!>\!=!})} to implement the \ensuremath{\Varid{allSix}} example, we lose the benefit of Curry-like non-determinism.
The row in \autoref{tab:allSix} labeled \enquote{Curry ND!} shows the running times when using \ensuremath{(\mathbin{>\!\!\!>\!\!\!>\!=!})} instead of \ensuremath{(\mathbin{>\!\!\!>\!\!\!>\!=})}.
As \ensuremath{(\mathbin{>\!\!\!>\!\!\!>\!=!})} is strict, the function \ensuremath{\Varid{joinWith}} has to evaluate both its arguments to yield a result.
\autoref{fig:evalstrict} shows how the formerly unneeded distribution \ensuremath{\Varid{dist}} now has to be evaluated in order to yield a value.
More precisely, using \ensuremath{(\mathbin{>\!\!\!>\!\!\!>\!=!})} causes a complete evaluation of \ensuremath{\Varid{dist}}.

\begin{figure}[t]
\centering
\begin{hscode}\SaveRestoreHook
\column{B}{@{}>{\hspre}l<{\hspost}@{}}%
\column{13}{@{}>{\hspre}l<{\hspost}@{}}%
\column{20}{@{}>{\hspre}l<{\hspost}@{}}%
\column{E}{@{}>{\hspre}l<{\hspost}@{}}%
\>[B]{}\Varid{filterDist}\;(\Varid{all}\;(\equiv \Conid{Six}))\;(\Varid{joinWith}\;(\mathbin{:})\;(\Conid{Dist}\;\Conid{One}\;\frac{\mathrm{1}}{\mathrm{6}})\;\Varid{dist}){}\<[E]%
\\
\>[B]{}\equiv \mbox{\commentbegin  Definition of \ensuremath{\Varid{joinWith}}  \commentend}{}\<[E]%
\\
\>[B]{}\Varid{filterDist}\;{}\<[13]%
\>[13]{}(\Varid{all}\;(\equiv \Conid{Six}))\;{}\<[E]%
\\
\>[13]{}(\Conid{Dist}\;\Conid{One}\;\frac{\mathrm{1}}{\mathrm{6}}\mathbin{>\!\!\!>\!\!\!>\!=!}(\lambda \Varid{x}\to \Varid{dist}\mathbin{>\!\!\!>\!\!\!>\!=!}(\lambda \Varid{xs}\to \Varid{certainly}\;(\Varid{x}\mathbin{:}\Varid{xs})))){}\<[E]%
\\
\>[B]{}\equiv \mbox{\commentbegin  Definition of \ensuremath{(\mathbin{>\!\!\!>\!\!\!>\!=!})}  \commentend}{}\<[E]%
\\
\>[B]{}\Varid{filterDist}\;{}\<[13]%
\>[13]{}(\Varid{all}\;(\equiv \Conid{Six}))\;{}\<[E]%
\\
\>[13]{}(\mathbf{case}\;{}\<[20]%
\>[20]{}(\lambda \Varid{x}\to \Varid{dist}\mathbin{>\!\!\!>\!\!\!>\!=!}(\lambda \Varid{xs}\to \Varid{certainly}\;(\Varid{x}\mathbin{:}\Varid{xs})))\;\Conid{One}\;\mathbf{of}{}\<[E]%
\\
\>[20]{}\Conid{Dist}\;\Varid{y}\;\Varid{q}\to \Conid{Dist}\;\Varid{y}\;(\frac{\mathrm{1}}{\mathrm{6}}\mathbin{*.}\Varid{q})){}\<[E]%
\\
\>[B]{}\equiv \mbox{\commentbegin  Evaluation of the scrutinee  \commentend}{}\<[E]%
\\
\>[B]{}\Varid{filterDist}\;{}\<[13]%
\>[13]{}(\Varid{all}\;(\equiv \Conid{Six}))\;{}\<[E]%
\\
\>[13]{}(\mathbf{case}\;{}\<[20]%
\>[20]{}\Varid{dist}\mathbin{>\!\!\!>\!\!\!>\!=!}(\lambda \Varid{xs}\to \Varid{certainly}\;(\Conid{One}\mathbin{:}\Varid{xs}))\;\mathbf{of}{}\<[E]%
\\
\>[20]{}\Conid{Dist}\;\Varid{y}\;\Varid{q}\to \Conid{Dist}\;\Varid{y}\;(\frac{\mathrm{1}}{\mathrm{6}}\mathbin{*.}\Varid{q})){}\<[E]%
\\
\>[B]{}\equiv \mbox{\commentbegin  Evaluation of \ensuremath{\Varid{dist}} as demanded by the definition of \ensuremath{(\mathbin{>\!\!\!>\!\!\!>\!=!})}  \commentend}{}\<[E]%
\\
\>[B]{}\mathbin{...}{}\<[E]%
\ColumnHook
\end{hscode}\resethooks
\caption{Simplified evaluation illustrating strict non-determinism}
\label{fig:evalstrict}
\end{figure}

Please note that an implementation that is similar to \ensuremath{(\mathbin{>\!\!\!>\!\!\!>\!=})} is \emph{not} possible in a list-based implementation.
The following definition of \ensuremath{\Varid{concatMap}} is usually used to define the bind operator for lists.
\begin{hscode}\SaveRestoreHook
\column{B}{@{}>{\hspre}l<{\hspost}@{}}%
\column{3}{@{}>{\hspre}l<{\hspost}@{}}%
\column{16}{@{}>{\hspre}l<{\hspost}@{}}%
\column{24}{@{}>{\hspre}l<{\hspost}@{}}%
\column{E}{@{}>{\hspre}l<{\hspost}@{}}%
\>[3]{}\Varid{concatMap}\mathbin{::}(\Varid{a}\to [\mskip1.5mu \Varid{b}\mskip1.5mu])\to [\mskip1.5mu \Varid{a}\mskip1.5mu]\to [\mskip1.5mu \Varid{b}\mskip1.5mu]{}\<[E]%
\\
\>[3]{}\Varid{concatMap}\;\Varid{f}\;{}\<[16]%
\>[16]{}[\mskip1.5mu \mskip1.5mu]{}\<[24]%
\>[24]{}\mathrel{=}[\mskip1.5mu \mskip1.5mu]{}\<[E]%
\\
\>[3]{}\Varid{concatMap}\;\Varid{f}\;{}\<[16]%
\>[16]{}(\Varid{x}\mathbin{:}\Varid{xs}){}\<[24]%
\>[24]{}\mathrel{=}\Varid{f}\;\Varid{x}\plus \Varid{concatMap}\;\Varid{f}\;\Varid{xs}{}\<[E]%
\ColumnHook
\end{hscode}\resethooks
The strict behavior follows from the definition via pattern matching on the list argument.
In contrast to \ensuremath{(\mathbin{>\!\!\!>\!\!\!>\!=!})} there is, however, no other implementation that is less strict.
The pattern matching is inevitable due to the two possible constructors, \ensuremath{[\mskip1.5mu \mskip1.5mu]} and \ensuremath{(\mathbin{:})}, for lists.
As a consequence, a list-based implementation has to traverse the entire distribution before we can evaluate the predicate \ensuremath{\Varid{all}\;(\equiv \Conid{Six})}.
The consequence is that the running times of \enquote{Haskell List} in \autoref{tab:allSix} cannot compete with \enquote{Curry ND} when the number of dice increases.

Intuitively, we expect similar running times for \enquote{Curry ND!} and \enquote{Curry List} as the bind operator for lists has to evaluate its second argument as well --- similar to \ensuremath{(\mathbin{>\!\!\!>\!\!\!>\!=!})}.
However, the observed running times do not have the expected resemblance.
\enquote{Curry ND!} heavily relies on non-deter\-mi\-nis\-tic computations, which causes significant overhead for KiCS2.
We do not investigate these differences here but propose it as a direction for future research.

Obviously, turning an exponential problem into a linear one is like getting only sixes when throwing dice.
In most cases we are not that lucky.
For example, consider the following query for throwing \ensuremath{\Varid{n}} dice that are either five or six.
\begin{hscode}\SaveRestoreHook
\column{B}{@{}>{\hspre}l<{\hspost}@{}}%
\column{3}{@{}>{\hspre}l<{\hspost}@{}}%
\column{E}{@{}>{\hspre}l<{\hspost}@{}}%
\>[3]{}\Varid{allFiveOrSix}\mathbin{::}\Conid{Int}\to \Conid{Probability}{}\<[E]%
\\
\>[3]{}\Varid{allFiveOrSix}\;\Varid{n}\mathrel{=}(\Varid{all}\;(\lambda \Varid{s}\to \Varid{s}\equiv \Conid{Five}\mathrel{\vee}\Varid{s}\equiv \Conid{Six}))\mathbin{??}(\Varid{replicateDist}\;\Varid{n}\;(\lambda ()\to \Varid{die})){}\<[E]%
\ColumnHook
\end{hscode}\resethooks

\autoref{tab:allFiveOrSix} lists the running times of this query for different numbers of dice with respect to the four different implementations.
\begin{table*}[h]\centering
\begin{tabular}{@{}lrrrrrr}
\toprule
\# of dice & 5 & 6 & 7 & 8 & 9 & 10\\
\midrule
Curry ND       & \num{4} & \num{7} & \num{15} & \num{34} & \num{76} & \num{163} \\
Curry List     & \num{2} & \num{13} & \num{84} & \num{489} & \num{2869} & \num{16989} \\
Curry ND!      & \num{49} & \num{382} & \num{2483} & \num{15562} & -- & -- \\
Haskell List   & \num{2} & \num{5} & \num{31} & \num{219} & \num{1423} & \num{6670} \\
\bottomrule
\end{tabular}
\caption{Overview of running times of the query \ensuremath{\Varid{allFiveOrSix}\;\Varid{n}}}
\label{tab:allFiveOrSix}
\end{table*}
As we can see from the running times, this query is exponential in all implementations.
Nevertheless, the running time of the non-strict, non-deterministic implementation is much better because we only have to consider two sides --- six and five --- while we have to consider all sides in the list implementations and the non-deterministic, strict implementation.
That is, while the base of the complexity is two in the case of the non-deterministic, non-strict implementation, it is six in all the other cases.
As we have observed in the other examples before, we get an overhead in the case of the strict non-determinism compared to the list implementation due to the heavy usage of non-deterministic computations.

\subsection{Definition of the Bind Operator}

In this subsection we discuss our design choices concerning the implementation of the bind operator.
We illustrate that we have to be careful about non-strictness, because we do not want to lose non-deterministic results.

First, we revisit the definition of \ensuremath{(\mathbin{>\!\!\!>\!\!\!>\!=})} introduced in~\autoref{sec:idea}.

\begin{hscode}\SaveRestoreHook
\column{B}{@{}>{\hspre}l<{\hspost}@{}}%
\column{13}{@{}>{\hspre}l<{\hspost}@{}}%
\column{18}{@{}>{\hspre}l<{\hspost}@{}}%
\column{E}{@{}>{\hspre}l<{\hspost}@{}}%
\>[B]{}(\mathbin{>\!\!\!>\!\!\!>\!=})\mathbin{::}\Conid{Dist}\;\Varid{a}\to (\Varid{a}\to \Conid{Dist}\;\Varid{b})\to \Conid{Dist}\;\Varid{b}{}\<[E]%
\\
\>[B]{}\Varid{d}\mathbin{>\!\!\!>\!\!\!>\!=}\Varid{f}\mathrel{=}{}\<[13]%
\>[13]{}\mathbf{let}\;{}\<[18]%
\>[18]{}\Conid{Dist}\;\Varid{x}\;\Varid{p}\mathrel{=}\Varid{d}{}\<[E]%
\\
\>[18]{}\Conid{Dist}\;\Varid{y}\;\Varid{q}\mathrel{=}\Varid{f}\;\Varid{x}{}\<[E]%
\\
\>[13]{}\mathbf{in}\;\Conid{Dist}\;\Varid{y}\;(\Varid{p}\mathbin{*.}\Varid{q}){}\<[E]%
\ColumnHook
\end{hscode}\resethooks
We can observe two facts about this definition.
First, the definition yields a \ensuremath{\Conid{Dist}}-constructor without matching any argument.
Second, if neither the event nor the probability of the final distribution is evaluated, the application of the function \ensuremath{\Varid{f}} is not evaluated either.

We can observe these properties with some exemplary usages of \ensuremath{(\mathbin{>\!\!\!>\!\!\!>\!=})}.
As a reference, we see that pattern matching the \ensuremath{\Conid{Dist}}-constructor of \ensuremath{\Varid{coin}} triggers the non-determinism and yields two results.
\begin{hscode}\SaveRestoreHook
\column{B}{@{}>{\hspre}l<{\hspost}@{}}%
\column{E}{@{}>{\hspre}l<{\hspost}@{}}%
\>[B]{}\mkern-3mu\mathbin{>}(\lambda (\Conid{Dist}\;\Varid{x}\;\Varid{p})\to \Conid{True})\;\Varid{coin}{}\<[E]%
\\
\>[B]{}\Conid{True}{}\<[E]%
\\
\>[B]{}\Conid{True}{}\<[E]%
\ColumnHook
\end{hscode}\resethooks
In contrast, distributions resulting from an application of \ensuremath{(\mathbin{>\!\!\!>\!\!\!>\!=})} behave differently.
This time, pattern matching on the \ensuremath{\Conid{Dist}}-constructor does not trigger any non-determinism.
\begin{hscode}\SaveRestoreHook
\column{B}{@{}>{\hspre}l<{\hspost}@{}}%
\column{E}{@{}>{\hspre}l<{\hspost}@{}}%
\>[B]{}\mkern-3mu\mathbin{>}(\lambda (\Conid{Dist}\;\Varid{x}\;\Varid{p})\to \Conid{True})\;(\Varid{certainly}\;()\mathbin{>\!\!\!>\!\!\!>\!=}(\lambda \Varid{y}\to \Varid{coin})){}\<[E]%
\\
\>[B]{}\Conid{True}{}\<[E]%
\\[\blanklineskip]%
\>[B]{}\mkern-3mu\mathbin{>}(\lambda (\Conid{Dist}\;\Varid{x}\;\Varid{p})\to \Conid{True})\;(\Varid{coin}\mathbin{>\!\!\!>\!\!\!>\!=}\Varid{certainly}){}\<[E]%
\\
\>[B]{}\Conid{True}{}\<[E]%
\ColumnHook
\end{hscode}\resethooks
We observe that the last two examples yield a single result, because the \ensuremath{(\mathbin{>\!\!\!>\!\!\!>\!=})}-operator changes the position of the non-determinism.
That is, the non-de\-ter\-min\-ism does not reside at the same level as the \ensuremath{\Conid{Dist}}-constructor, but in the arguments of \ensuremath{\Conid{Dist}}.
Therefore, we have to be sure to trigger all non-determinism when we query distributions.
Not evaluating non-determinism might lead to false results when we sum up probabilities.
Hence, non-strictness is a crucial property for positive pruning effects, but has to be used carefully.

Consider the following example usage of \ensuremath{(\mathbin{>\!\!\!>\!\!\!>\!=})}, which is an inlined version of \ensuremath{\Varid{joinWith}} applied to the Boolean conjunction \ensuremath{(\mathrel{\wedge})}.

\begin{hscode}\SaveRestoreHook
\column{B}{@{}>{\hspre}l<{\hspost}@{}}%
\column{E}{@{}>{\hspre}l<{\hspost}@{}}%
\>[B]{}\mkern-3mu\mathbin{>}(\lambda (\Conid{Dist}\;\Varid{x}\;\Varid{p})\to \Varid{x})\;(\Varid{coin}\mathbin{>\!\!\!>\!\!\!>\!=}(\lambda \Varid{b1}\to \Varid{coin}\mathbin{>\!\!\!>\!\!\!>\!=}(\lambda \Varid{b2}\to \Varid{certainly}\;(\Varid{b1}\mathrel{\wedge}\Varid{b2})))){}\<[E]%
\\
\>[B]{}\Conid{False}{}\<[E]%
\\
\>[B]{}\Conid{True}{}\<[E]%
\\
\>[B]{}\Conid{False}{}\<[E]%
\ColumnHook
\end{hscode}\resethooks
We lose one expected result from the distribution, because \ensuremath{(\mathrel{\wedge})} is non-strict in its second argument in case the first argument is \ensuremath{\Conid{False}}.
When the first \ensuremath{\Varid{coin}} evaluates to \ensuremath{\Conid{False}}, \ensuremath{(\mathbin{>\!\!\!>\!\!\!>\!=})} ignores the second coin and yields \ensuremath{\Conid{False}} straightaway.
In this case, the non-determinism of the second \ensuremath{\Varid{coin}} is not triggered and we get only three instead of four results.
The non-strictness of \ensuremath{(\mathrel{\wedge})} has no consequences when using \ensuremath{(\mathbin{>\!\!\!>\!\!\!>\!=!})}, because the operator evaluates both arguments and, thus, triggers the non-determinism.
In the case of projecting to the event we do not care about the missing result.
However, when we sum up probabilities, we do not want events to get lost.

When we compute the total probability of a distribution, the result should always be \ensuremath{\mathrm{1.0}}.
However, the query above has only three results and every event has a probability of \ensuremath{\mathrm{0.25}}, resulting in a total probability of \ensuremath{\mathrm{0.75}}.
Here is the good news.
While events can get lost when passing non-strict functions to \ensuremath{(\mathbin{>\!\!\!>\!\!\!>\!=})}, probabilities never get lost.
For example, consider the following application.

\begin{hscode}\SaveRestoreHook
\column{B}{@{}>{\hspre}l<{\hspost}@{}}%
\column{E}{@{}>{\hspre}l<{\hspost}@{}}%
\>[B]{}\mkern-3mu\mathbin{>}(\lambda (\Conid{Dist}\;\Varid{x}\;\Varid{p})\to \Varid{p})\;(\Varid{coin}\mathbin{>\!\!\!>\!\!\!>\!=}(\lambda \Varid{b1}\to \Varid{coin}\mathbin{>\!\!\!>\!\!\!>\!=}(\lambda \Varid{b2}\to \Varid{certainly}\;(\Varid{b1}\mathrel{\wedge}\Varid{b2})))){}\<[E]%
\\
\>[B]{}\mathrm{0.25}{}\<[E]%
\\
\>[B]{}\mathrm{0.25}{}\<[E]%
\\
\>[B]{}\mathrm{0.25}{}\<[E]%
\\
\>[B]{}\mathrm{0.25}{}\<[E]%
\ColumnHook
\end{hscode}\resethooks

Since multiplication is strict, if we demand the resulting probability, the operator \ensuremath{(\mathbin{>\!\!\!>\!\!\!>\!=})} has to evaluate the \ensuremath{\Conid{Dist}}-constructor and its probability.
That is, no values get lost if we evaluate the resulting probability.
Fortunately, the query operation \ensuremath{(\mathbin{??})} calculates the total probability of the filtered distributions, thus, evaluates the probability as the following example shows.

\begin{hscode}\SaveRestoreHook
\column{B}{@{}>{\hspre}l<{\hspost}@{}}%
\column{E}{@{}>{\hspre}l<{\hspost}@{}}%
\>[B]{}\mkern-3mu\mathbin{>}not\mathbin{??}(\Varid{coin}\mathbin{>\!\!\!>\!\!\!>\!=}(\lambda \Varid{b1}\to \Varid{coin}\mathbin{>\!\!\!>\!\!\!>\!=}(\lambda \Varid{b2}\to \Varid{certainly}\;(\Varid{b1}\mathrel{\wedge}\Varid{b2})))){}\<[E]%
\\
\>[B]{}\mathrm{0.75}{}\<[E]%
\ColumnHook
\end{hscode}\resethooks
We calculate the probability of the event \ensuremath{\Conid{False}}.
While there were only two \ensuremath{\Conid{False}} events when we projected to the event, the total probability of the event \ensuremath{\Conid{False}} is still \ensuremath{\mathrm{0.75}}, i.e., three times \ensuremath{\mathrm{0.25}}, instead of only \ensuremath{\mathrm{0.5}}.

All in all, in order to benefit from non-strictness, all operations provided by the library have to use the right amount of strictness, not too much and not too little.
For this reason PFLP does not provide the \ensuremath{\Conid{Dist}}-constructor nor the corresponding projection functions to the user.
With this restriction, the library guarantees that no relevant probabilities get lost.

\subsection{Non-deterministic Events}

We assume that all events passed to library functions are deterministic, that is, the library does not support non-deterministic events within distributions.
In order to illustrate why this restriction is crucial, we consider an example that breaks this rule.

Curry provides free variables, that is, expressions that non-deterministically evaluate to every possible value of its type.
When we revisit the definition of a die, we might be tempted to use a free variable instead of explicitly enumerating all values of type \ensuremath{\Conid{Side}}.

We can define a free variable of type \ensuremath{\Conid{Side}} as follows.
\begin{hscode}\SaveRestoreHook
\column{B}{@{}>{\hspre}l<{\hspost}@{}}%
\column{3}{@{}>{\hspre}l<{\hspost}@{}}%
\column{E}{@{}>{\hspre}l<{\hspost}@{}}%
\>[3]{}\Varid{side}\mathbin{::}\Conid{Side}{}\<[E]%
\\
\>[3]{}\Varid{side}\mathrel{=}\Varid{unknown}{}\<[E]%
\ColumnHook
\end{hscode}\resethooks
This free variable evaluates as follows.
\begin{hscode}\SaveRestoreHook
\column{B}{@{}>{\hspre}l<{\hspost}@{}}%
\column{E}{@{}>{\hspre}l<{\hspost}@{}}%
\>[B]{}\mkern-3mu\mathbin{>}\Varid{side}{}\<[E]%
\\
\>[B]{}\Conid{One}{}\<[E]%
\\
\>[B]{}\Conid{Two}{}\<[E]%
\\
\>[B]{}\Conid{Three}{}\<[E]%
\\
\>[B]{}\Conid{Four}{}\<[E]%
\\
\>[B]{}\Conid{Five}{}\<[E]%
\\
\>[B]{}\Conid{Six}{}\<[E]%
\ColumnHook
\end{hscode}\resethooks
With this information in mind consider the following alternative definition of a die, which is much more concise than explicitly listing all constructors of \ensuremath{\Conid{Dist}}.
\begin{hscode}\SaveRestoreHook
\column{B}{@{}>{\hspre}l<{\hspost}@{}}%
\column{3}{@{}>{\hspre}l<{\hspost}@{}}%
\column{E}{@{}>{\hspre}l<{\hspost}@{}}%
\>[3]{}\Varid{die2}\mathbin{::}\Conid{Dist}\;\Conid{Side}{}\<[E]%
\\
\>[3]{}\Varid{die2}\mathrel{=}\Varid{enum}\;[\mskip1.5mu \Varid{side}\mskip1.5mu]\;[\mskip1.5mu \frac{\mathrm{1}}{\mathrm{6}}\mskip1.5mu]{}\<[E]%
\ColumnHook
\end{hscode}\resethooks
We just use a free variable --- the constant \ensuremath{\Varid{side}} --- and pass the probability of each event as second parameter.
Now, let us consider the following query.

\begin{hscode}\SaveRestoreHook
\column{B}{@{}>{\hspre}l<{\hspost}@{}}%
\column{E}{@{}>{\hspre}l<{\hspost}@{}}%
\>[B]{}\mkern-3mu\mathbin{>}(\Varid{const}\;\Conid{True})\mathbin{??}\Varid{die2}{}\<[E]%
\\
\>[B]{}\mathrm{0.16666667}{}\<[E]%
\ColumnHook
\end{hscode}\resethooks

The result of this query is \ensuremath{\frac{\mathrm{1}}{\mathrm{6}}} and not \ensuremath{\mathrm{1.0}} as expected.
Consider \autoref{fig:unknown} for a step-by-step evaluation of this expression in order to understand better what is going on.
This example illustrates that probabilities can get lost if we do not use the right amount of strictness.
The predicate \ensuremath{\Varid{const}\;\Conid{True}} does not touch the event at all, thus does not trigger \ensuremath{\Varid{side}} to actually evaluate to all the constructors of \ensuremath{\Conid{Side}}.
Then, the definition of \ensuremath{(\mathbin{??})} directly projects to the probability of \ensuremath{\Varid{die2}} and throws away all non-determinism left in \ensuremath{\Conid{Dist}\;\Varid{side}\;\frac{\mathrm{1}}{\mathrm{6}}}.
Therefore, we lose probabilities we would like to sum up.

\begin{figure}
\begin{hscode}\SaveRestoreHook
\column{B}{@{}>{\hspre}l<{\hspost}@{}}%
\column{E}{@{}>{\hspre}l<{\hspost}@{}}%
\>[B]{}(\Varid{const}\;\Conid{True})\mathbin{??}\Varid{die2}{}\<[E]%
\\
\>[B]{}\equiv \mbox{\commentbegin  definition of \ensuremath{(\mathbin{??})}  \commentend}{}\<[E]%
\\
\>[B]{}\Varid{foldValues}\;(\mathbin{+.})\;\mathrm{0.0}\;(\Varid{allValues}\;(\Varid{prob}\;(\Varid{filterDist}\;(\Varid{const}\;\Conid{True})\;\Varid{die2}))){}\<[E]%
\\
\>[B]{}\equiv \mbox{\commentbegin  definition of \ensuremath{\Varid{die2}}  \commentend}{}\<[E]%
\\
\>[B]{}\Varid{foldValues}\;(\mathbin{+.})\;\mathrm{0.0}\;(\Varid{allValues}\;(\Varid{prob}\;(\Varid{filterDist}\;(\Varid{const}\;\Conid{True})\;(\Varid{enum}\;[\mskip1.5mu \Varid{side}\mskip1.5mu]\;[\mskip1.5mu \frac{\mathrm{1}}{\mathrm{6}}\mskip1.5mu])))){}\<[E]%
\\
\>[B]{}\equiv \mbox{\commentbegin  definition of \ensuremath{\Varid{enum}}  \commentend}{}\<[E]%
\\
\>[B]{}\Varid{foldValues}\;(\mathbin{+.})\;\mathrm{0.0}\;(\Varid{allValues}\;(\Varid{prob}\;(\Varid{filterDist}\;(\Varid{const}\;\Conid{True})\;(\Conid{Dist}\;\Varid{side}\;\frac{\mathrm{1}}{\mathrm{6}})))){}\<[E]%
\\
\>[B]{}\equiv \mbox{\commentbegin  definition of \ensuremath{\Varid{filterDist}}  \commentend}{}\<[E]%
\\
\>[B]{}\Varid{foldValues}\;(\mathbin{+.})\;\mathrm{0.0}\;(\Varid{allValues}\;(\Varid{prob}\;(\mathbf{if}\;\Varid{const}\;\Conid{True}\;\Varid{side}\;\mathbf{then}\;\Conid{Dist}\;\Varid{side}\;\frac{\mathrm{1}}{\mathrm{6}}\;\mathbf{else}\;\Varid{failed}))){}\<[E]%
\\
\>[B]{}\equiv \mbox{\commentbegin  definition of \ensuremath{\Varid{const}}  \commentend}{}\<[E]%
\\
\>[B]{}\Varid{foldValues}\;(\mathbin{+.})\;\mathrm{0.0}\;(\Varid{allValues}\;(\Varid{prob}\;(\mathbf{if}\;\Conid{True}\;\mathbf{then}\;\Conid{Dist}\;\Varid{side}\;\frac{\mathrm{1}}{\mathrm{6}}\;\mathbf{else}\;\Varid{failed}))){}\<[E]%
\\
\>[B]{}\equiv \mbox{\commentbegin  evaluate \ensuremath{\mathbf{if}\mathbin{-}\mathbf{then}\mathbin{-}\mathbf{else}}  \commentend}{}\<[E]%
\\
\>[B]{}\Varid{foldValues}\;(\mathbin{+.})\;\mathrm{0.0}\;(\Varid{allValues}\;(\Varid{prob}\;(\Conid{Dist}\;\Varid{side}\;\frac{\mathrm{1}}{\mathrm{6}}))){}\<[E]%
\\
\>[B]{}\equiv \mbox{\commentbegin  definition of \ensuremath{\Varid{prob}}  \commentend}{}\<[E]%
\\
\>[B]{}\Varid{foldValues}\;(\mathbin{+.})\;\mathrm{0.0}\;(\Varid{allValues}\;\frac{\mathrm{1}}{\mathrm{6}}){}\<[E]%
\\
\>[B]{}\equiv \mbox{\commentbegin  definition of \ensuremath{\Varid{allValues}}  \commentend}{}\<[E]%
\\
\>[B]{}\Varid{foldValues}\;(\mathbin{+.})\;\mathrm{0.0}\;\{\mskip1.5mu \frac{\mathrm{1}}{\mathrm{6}}\mskip1.5mu\}{}\<[E]%
\\
\>[B]{}\equiv \mbox{\commentbegin  definition of \ensuremath{\Varid{foldValues}}  \commentend}{}\<[E]%
\\
\>[B]{}\frac{\mathrm{1}}{\mathrm{6}}{}\<[E]%
\ColumnHook
\end{hscode}\resethooks
\caption{Evaluation of a distribution that contains a free variable that is not demanded}
\label{fig:unknown}
\end{figure}

As a consequence for PFLP, non-deterministic events within a distribution are not allowed.
If users of the library stick to this rule, it is not possible to misuse the operations and lose non-deterministic results due to non-strictness.

However, one approach to overcome this issue when using \ensuremath{\Varid{enum}} is to use an alternative stricter implementation.
That is, we could easily adapt the strictness behavior of \ensuremath{\Varid{enum}} in order to allow a more declarative definition of distributions using free variables without affecting the overall advantage leveraged by non-strict functions.

\subsection{Partial Functions}

Besides not using non-determinism for events, users have to follow another restriction.
When using the bind operator \ensuremath{(\mathbin{>\!\!\!>\!\!\!>\!=})}, the second argument is a function of type \ensuremath{\Varid{a}\to \Conid{Dist}\;\Varid{b}}, that is, constructs a new distribution.
As we have discussed before distributions need to sum up to a probability of $1.0$, and the distributions we create via \ensuremath{(\mathbin{>\!\!\!>\!\!\!>\!=})} are no exception.
This restriction is violated if we use partial functions as second argument of \ensuremath{(\mathbin{>\!\!\!>\!\!\!>\!=})}.
Recall the definition \ensuremath{\Varid{coin}} that describes a uniform distribution of type \ensuremath{\Conid{Bool}}, and consider the function \ensuremath{\Varid{partialPattern}} that depends on \ensuremath{\Varid{coin}}, but maps \ensuremath{\Conid{False}} to \ensuremath{\Varid{failed}}.
\begin{hscode}\SaveRestoreHook
\column{B}{@{}>{\hspre}l<{\hspost}@{}}%
\column{3}{@{}>{\hspre}l<{\hspost}@{}}%
\column{E}{@{}>{\hspre}l<{\hspost}@{}}%
\>[3]{}\Varid{partialPattern}\mathbin{::}\Conid{Dist}\;\Conid{Bool}{}\<[E]%
\\
\>[3]{}\Varid{partialPattern}\mathrel{=}\Varid{coin}\mathbin{>\!\!\!>\!\!\!>\!=}(\lambda \Varid{b}\to \mathbf{if}\;\Varid{b}\;\mathbf{then}\;\Varid{certainly}\;\Conid{True}\;\mathbf{else}\;\Varid{failed}){}\<[E]%
\ColumnHook
\end{hscode}\resethooks
Due to the partial pattern matching in \ensuremath{\Varid{partialPattern}}, the resulting distribution does not sum up to $1.0$ anymore, thus, violates the rule for a valid distribution.
By performing a query with the predicate \ensuremath{\Varid{const}\;\Conid{True}} we can observe this property.

\begin{hscode}\SaveRestoreHook
\column{B}{@{}>{\hspre}l<{\hspost}@{}}%
\column{E}{@{}>{\hspre}l<{\hspost}@{}}%
\>[B]{}\mkern-3mu\mathbin{>}(\Varid{const}\;\Conid{True})\mathbin{??}\Varid{partialPattern}{}\<[E]%
\\
\>[B]{}\mathrm{0.5}{}\<[E]%
\ColumnHook
\end{hscode}\resethooks

We only allow to filter distributions when a probability is computed using \ensuremath{(\mathbin{??})}, but not in any other situation.
In the current implementation this restriction on functions when using \ensuremath{(\mathbin{>\!\!\!>\!\!\!>\!=})} is neither statically nor dynamically enforced, but a coding convention that users should keep in mind and follow when working with the library.

\subsection{Monad Laws}
\label{ssec:monad}

When we comply with the restrictions we have discussed above, the operators \ensuremath{(\mathbin{>\!\!\!>\!\!\!>\!=})} and \ensuremath{\Varid{certainly}} allow us to formulate probabilistic programs as one would expect.
However, there is one obvious question that we did not answer yet.
We did not check whether the operator \ensuremath{(\mathbin{>\!\!\!>\!\!\!>\!=})} together with \ensuremath{\Varid{certainly}} actually forms a monad as the name of the operator suggests.
That is, we have to check whether the following three laws hold for all distributions \ensuremath{\Varid{d}} and all values \ensuremath{\Varid{x}}, \ensuremath{\Varid{f}}, and \ensuremath{\Varid{g}} of appropriate types.

\begin{itemize}
\item \ensuremath{\Varid{d}\mathbin{>\!\!\!>\!\!\!>\!=}\Varid{certainly}\equiv \Varid{d}}
\item \ensuremath{\Varid{certainly}\;\Varid{x}\mathbin{>\!\!\!>\!\!\!>\!=}\Varid{f}\equiv \Varid{f}\;\Varid{x}}
\item \ensuremath{(\Varid{d}\mathbin{>\!\!\!>\!\!\!>\!=}\Varid{f})\mathbin{>\!\!\!>\!\!\!>\!=}\Varid{g}\equiv \Varid{d}\mathbin{>\!\!\!>\!\!\!>\!=}(\lambda \Varid{y}\to \Varid{f}\;\Varid{y}\mathbin{>\!\!\!>\!\!\!>\!=}\Varid{g})}
\end{itemize}

In the previous subsection we have already observed that the first equality does not hold in general.
For example, we have seen that there is a context that is able to distinguish the left-hand from the right-hand side.
For instance, while the expression
\[
\ensuremath{(\lambda (\Conid{Dist}\;\Varid{x}\;\Varid{p})\to \Conid{True})\;\Varid{coin}}
\]
yields \ensuremath{\Conid{True}} twice, the expression
\[
\ensuremath{(\lambda (\Conid{Dist}\;\Varid{x}\;\Varid{p})\to \Conid{True})\;(\Varid{coin}\mathbin{>\!\!\!>\!\!\!>\!=}\Varid{certainly})}
\]
yields \ensuremath{\Conid{True}} only once.
As most Curry semantics are based on sets --- and not on multisets, the two sides of the equality would be the same.
Notwithstanding, in a multiset semantics the user could still not observe the difference between the two expressions because he does not have access to the \ensuremath{\Conid{Dist}}-constructor.
The user cannot pattern match on a \ensuremath{\Conid{Dist}}-constructor, but only use the combinator \ensuremath{(\mathbin{??})} to inspect a distribution.

In order to discuss the validity of the monad laws more rigorously, we apply equational reasoning to check whether the monad laws might fail.

\paragraph{The first monad law}
Let \ensuremath{\Varid{d}\mathbin{::}\Conid{Dist}\;\tau } then we reason as follows about the first monad law.

\begin{hscode}\SaveRestoreHook
\column{B}{@{}>{\hspre}l<{\hspost}@{}}%
\column{6}{@{}>{\hspre}l<{\hspost}@{}}%
\column{E}{@{}>{\hspre}l<{\hspost}@{}}%
\>[B]{}\Varid{d}\mathbin{>\!\!\!>\!\!\!>\!=}\Varid{certainly}{}\<[E]%
\\
\>[B]{}\equiv \mbox{\commentbegin  Definition of \ensuremath{(\mathbin{>\!\!\!>\!\!\!>\!=})}  \commentend}{}\<[E]%
\\
\>[B]{}\mathbf{let}\;{}\<[6]%
\>[6]{}\Conid{Dist}\;\Varid{x}\;\Varid{p}\mathrel{=}\Varid{d}{}\<[E]%
\\
\>[6]{}\Conid{Dist}\;\Varid{y}\;\Varid{q}\mathrel{=}\Varid{certainly}\;\Varid{x}{}\<[E]%
\\
\>[B]{}\mathbf{in}\;\Conid{Dist}\;\Varid{y}\;(\Varid{p}\mathbin{*.}\Varid{q}){}\<[E]%
\\
\>[B]{}\equiv \mbox{\commentbegin  Definition of \ensuremath{\Varid{certainly}}  \commentend}{}\<[E]%
\\
\>[B]{}\mathbf{let}\;{}\<[6]%
\>[6]{}\Conid{Dist}\;\Varid{x}\;\Varid{p}\mathrel{=}\Varid{d}{}\<[E]%
\\
\>[6]{}\Conid{Dist}\;\Varid{y}\;\Varid{q}\mathrel{=}\Conid{Dist}\;\Varid{x}\;\mathrm{1.0}{}\<[E]%
\\
\>[B]{}\mathbf{in}\;\Conid{Dist}\;\Varid{y}\;(\Varid{p}\mathbin{*.}\Varid{q}){}\<[E]%
\\
\>[B]{}\equiv \mbox{\commentbegin  Inlining of \ensuremath{\Conid{Dist}\;\Varid{y}\;\Varid{q}\mathrel{=}\Conid{Dist}\;\Varid{x}\;\mathrm{1.0}}  \commentend}{}\<[E]%
\\
\>[B]{}\mathbf{let}\;\Conid{Dist}\;\Varid{x}\;\Varid{p}\mathrel{=}\Varid{d}\;\mathbf{in}\;\Conid{Dist}\;\Varid{x}\;(\Varid{p}\mathbin{*.}\mathrm{1.0}){}\<[E]%
\\
\>[B]{}\equiv \mbox{\commentbegin  Definition of \ensuremath{(\mathbin{*.})}  \commentend}{}\<[E]%
\\
\>[B]{}\mathbf{let}\;\Conid{Dist}\;\Varid{x}\;\Varid{p}\mathrel{=}\Varid{d}\;\mathbf{in}\;\Conid{Dist}\;\Varid{x}\;\Varid{p}{}\<[E]%
\\
\>[B]{}\stackrel{?}{\equiv}{}\<[E]%
\\
\>[B]{}\Varid{d}{}\<[E]%
\ColumnHook
\end{hscode}\resethooks
Does the last step hold in general? It looks good for the deterministic case with \ensuremath{\Varid{d}\mathrel{=}\Conid{Dist}\;\Varid{evnt}\;\Varid{prb}}.

\begin{hscode}\SaveRestoreHook
\column{B}{@{}>{\hspre}l<{\hspost}@{}}%
\column{E}{@{}>{\hspre}l<{\hspost}@{}}%
\>[B]{}\mathbf{let}\;\Conid{Dist}\;\Varid{x}\;\Varid{p}\mathrel{=}\Conid{Dist}\;\Varid{evnt}\;\Varid{prb}\;\mathbf{in}\;\Conid{Dist}\;\Varid{x}\;\Varid{p}{}\<[E]%
\\
\>[B]{}\equiv {}\<[E]%
\\
\>[B]{}\Conid{Dist}\;\Varid{evnt}\;\Varid{prb}{}\<[E]%
\ColumnHook
\end{hscode}\resethooks
However, the equality \ensuremath{\mathbf{let}\;\Conid{Dist}\;\Varid{x}\;\Varid{p}\mathrel{=}\Varid{d}\;\mathbf{in}\;\Conid{Dist}\;\Varid{x}\;\Varid{p}\equiv \Varid{d}} does not hold in general.
For instance let us consider the case \ensuremath{\Varid{d}\mathrel{=}\Varid{failed}}.
\begin{hscode}\SaveRestoreHook
\column{B}{@{}>{\hspre}l<{\hspost}@{}}%
\column{E}{@{}>{\hspre}l<{\hspost}@{}}%
\>[B]{}\mathbf{let}\;\Conid{Dist}\;\Varid{x}\;\Varid{p}\mathrel{=}\Varid{failed}\;\mathbf{in}\;\Conid{Dist}\;\Varid{x}\;\Varid{p}{}\<[E]%
\\
\>[B]{}\equiv {}\<[E]%
\\
\>[B]{}\Conid{Dist}\;\Varid{failed}\;\Varid{failed}{}\<[E]%
\\
\>[B]{}\not\equiv {}\<[E]%
\\
\>[B]{}\Varid{failed}{}\<[E]%
\ColumnHook
\end{hscode}\resethooks
That is, the left-hand side is more defined than the right-hand side if \ensuremath{\Varid{d}\mathrel{=}\Varid{failed}}.

Because the user cannot access the \ensuremath{\Conid{Dist}}-constructor she cannot observe this difference.
The user can only compare two distributions by using the querying operator \ensuremath{(\mathbin{??})}.
Therefore, in the following we will show that the monad laws hold if we consider a context of the form \ensuremath{\Varid{pred}\mathbin{??}\Varid{d}} where \ensuremath{\Varid{pred}} is an arbitrary predicate.
Recall that we defined the operator \ensuremath{(\mathbin{??})} as follows.

\begin{hscode}\SaveRestoreHook
\column{B}{@{}>{\hspre}l<{\hspost}@{}}%
\column{E}{@{}>{\hspre}l<{\hspost}@{}}%
\>[B]{}(\mathbin{??})\mathbin{::}(\Varid{a}\to \Conid{Bool})\to \Conid{Dist}\;\Varid{a}\to \Conid{Probability}{}\<[E]%
\\
\>[B]{}(\mathbin{??})\;\Varid{pred}\;\Varid{d}\mathrel{=}\Varid{foldValues}\;(\mathbin{+.})\;\mathrm{0.0}\;(\Varid{allValues}\;(\Varid{prob}\;(\Varid{filterDist}\;\Varid{pred}\;\Varid{d}))){}\<[E]%
\ColumnHook
\end{hscode}\resethooks

Fortunately, the monad laws already hold if we consider the context \ensuremath{\Varid{filterDist}\;\Varid{pred}} for an arbitrary predicate \ensuremath{\Varid{pred}\mathbin{::}\Varid{a}\to \Conid{Bool}}.
Therefore we will show that the following equalities hold for all distributions \ensuremath{\Varid{d}}, and all values \ensuremath{\Varid{x}}, \ensuremath{\Varid{pred}}, \ensuremath{\Varid{f}}, and \ensuremath{\Varid{g}} of appropriate types.

\begin{enumerate}[ref=(\arabic*),label=(\arabic*)]
\item \label{item:law1} \ensuremath{\Varid{filterDist}\;\Varid{pred}\;(\Varid{d}\mathbin{>\!\!\!>\!\!\!>\!=}\Varid{certainly})\equiv \Varid{filterDist}\;\Varid{pred}\;\Varid{d}}
\item \label{item:law2} \ensuremath{\Varid{filterDist}\;\Varid{pred}\;(\Varid{certainly}\;\Varid{x}\mathbin{>\!\!\!>\!\!\!>\!=}\Varid{f})\equiv \Varid{filterDist}\;\Varid{pred}\;(\Varid{f}\;\Varid{x})}
\item \label{item:law3} \ensuremath{\Varid{filterDist}\;\Varid{pred}\;((\Varid{d}\mathbin{>\!\!\!>\!\!\!>\!=}\Varid{f})\mathbin{>\!\!\!>\!\!\!>\!=}\Varid{g}))\equiv \Varid{filterDist}\;\Varid{pred}\;(\Varid{d}\mathbin{>\!\!\!>\!\!\!>\!=}(\lambda \Varid{y}\to \Varid{f}\;\Varid{y}\mathbin{>\!\!\!>\!\!\!>\!=}\Varid{g}))}
\end{enumerate}

In the following we will first show that equation \ref{item:law1} holds.
We reason as follows for all distributions \ensuremath{\Varid{d}\mathbin{::}\Conid{Dist}\;\tau } and predicates \ensuremath{\Varid{pred}\mathbin{::}\tau \to \Conid{Bool}}.

\begin{hscode}\SaveRestoreHook
\column{B}{@{}>{\hspre}l<{\hspost}@{}}%
\column{E}{@{}>{\hspre}l<{\hspost}@{}}%
\>[B]{}\Varid{filterDist}\;\Varid{pred}\;(\Varid{d}\mathbin{>\!\!\!>\!\!\!>\!=}\Varid{certainly}){}\<[E]%
\\
\>[B]{}\equiv \mbox{\commentbegin  Reasoning above  \commentend}{}\<[E]%
\\
\>[B]{}\Varid{filterDist}\;\Varid{pred}\;(\mathbf{let}\;\Conid{Dist}\;\Varid{x}\;\Varid{p}\mathrel{=}\Varid{d}\;\mathbf{in}\;\Conid{Dist}\;\Varid{x}\;\Varid{p}){}\<[E]%
\\
\>[B]{}\equiv \mbox{\commentbegin  Definition of \ensuremath{\Varid{filterDist}}  \commentend}{}\<[E]%
\\
\>[B]{}\mathbf{let}\;\Conid{Dist}\;\Varid{y}\;\Varid{q}\mathrel{=}(\mathbf{let}\;\Conid{Dist}\;\Varid{x}\;\Varid{p}\mathrel{=}\Varid{d}\;\mathbf{in}\;\Conid{Dist}\;\Varid{x}\;\Varid{p}){}\<[E]%
\\
\>[B]{}\mathbf{in}\;\mathbf{if}\;(\Varid{pred}\;\Varid{y})\;\mathbf{then}\;(\Conid{Dist}\;\Varid{y}\;\Varid{q})\;\mathbf{else}\;\Varid{failed}{}\<[E]%
\\
\>[B]{}\equiv \mbox{\commentbegin  Inline \ensuremath{\mathbf{let}}-declaration  \commentend}{}\<[E]%
\\
\>[B]{}\mathbf{let}\;\Conid{Dist}\;\Varid{y}\;\Varid{q}\mathrel{=}\Varid{d}{}\<[E]%
\\
\>[B]{}\mathbf{in}\;\mathbf{if}\;(\Varid{pred}\;\Varid{y})\;\mathbf{then}\;(\Conid{Dist}\;\Varid{y}\;\Varid{q})\;\mathbf{else}\;\Varid{failed}{}\<[E]%
\\
\>[B]{}\equiv \mbox{\commentbegin  Definition of \ensuremath{\Varid{filterDist}}  \commentend}{}\<[E]%
\\
\>[B]{}\Varid{filterDist}\;\Varid{pred}\;\Varid{d}{}\<[E]%
\ColumnHook
\end{hscode}\resethooks

The \ensuremath{(\mathbin{>\!\!\!>\!\!\!>\!=})}-operator defers the pattern matching to the right-hand side via a \ensuremath{\mathbf{let}}-expression.
This so-called lazy pattern matching causes the monad laws to not hold without any context.
However, because \ensuremath{\Varid{filterDist}} introduces a lazy pattern matching via a \ensuremath{\mathbf{let}}-expression as well, observing two distributions via \ensuremath{\Varid{filterDist}} hides the difference between the two sides of the equation.

\paragraph{The second monad law}
For the second monad law \ref{item:law2} we reason as follows for all \ensuremath{\Varid{x}\mathbin{::}\tau _{1}}, and all \ensuremath{\Varid{f}\mathbin{::}\tau _{1}\to \Conid{Dist}\;\tau _{2}}.
\begin{hscode}\SaveRestoreHook
\column{B}{@{}>{\hspre}l<{\hspost}@{}}%
\column{6}{@{}>{\hspre}l<{\hspost}@{}}%
\column{E}{@{}>{\hspre}l<{\hspost}@{}}%
\>[B]{}\Varid{certainly}\;\Varid{x}\mathbin{>\!\!\!>\!\!\!>\!=}\Varid{f}{}\<[E]%
\\
\>[B]{}\equiv \mbox{\commentbegin  Definition of \ensuremath{(\mathbin{>\!\!\!>\!\!\!>\!=})}  \commentend}{}\<[E]%
\\
\>[B]{}\mathbf{let}\;{}\<[6]%
\>[6]{}\Conid{Dist}\;\Varid{y}\;\Varid{p}\mathrel{=}\Varid{certainly}\;\Varid{x}{}\<[E]%
\\
\>[6]{}\Conid{Dist}\;\Varid{z}\;\Varid{q}\mathrel{=}\Varid{f}\;\Varid{y}{}\<[E]%
\\
\>[B]{}\mathbf{in}\;\Conid{Dist}\;\Varid{z}\;(\Varid{p}\mathbin{*.}\Varid{q}){}\<[E]%
\\
\>[B]{}\equiv \mbox{\commentbegin  Definition of \ensuremath{\Varid{certainly}}  \commentend}{}\<[E]%
\\
\>[B]{}\mathbf{let}\;{}\<[6]%
\>[6]{}\Conid{Dist}\;\Varid{y}\;\Varid{p}\mathrel{=}\Conid{Dist}\;\Varid{x}\;\mathrm{1.0}{}\<[E]%
\\
\>[6]{}\Conid{Dist}\;\Varid{z}\;\Varid{q}\mathrel{=}\Varid{f}\;\Varid{y}{}\<[E]%
\\
\>[B]{}\mathbf{in}\;\Conid{Dist}\;\Varid{z}\;(\Varid{p}\mathbin{*.}\Varid{q}){}\<[E]%
\\
\>[B]{}\equiv \mbox{\commentbegin  Inlining of \ensuremath{\Conid{Dist}\;\Varid{y}\;\Varid{p}\mathrel{=}\Conid{Dist}\;\Varid{x}\;\mathrm{1.0}}  \commentend}{}\<[E]%
\\
\>[B]{}\mathbf{let}\;\Conid{Dist}\;\Varid{z}\;\Varid{q}\mathrel{=}\Varid{f}\;\Varid{x}\;\mathbf{in}\;\Conid{Dist}\;\Varid{z}\;(\mathrm{1.0}\mathbin{*.}\Varid{q}){}\<[E]%
\\
\>[B]{}\equiv \mbox{\commentbegin  Definition of \ensuremath{(\mathbin{*.})}  \commentend}{}\<[E]%
\\
\>[B]{}\mathbf{let}\;\Conid{Dist}\;\Varid{z}\;\Varid{q}\mathrel{=}\Varid{f}\;\Varid{x}\;\mathbf{in}\;\Conid{Dist}\;\Varid{z}\;\Varid{q}{}\<[E]%
\\
\>[B]{}\stackrel{?}{\equiv}{}\<[E]%
\\
\>[B]{}\Varid{f}\;\Varid{x}{}\<[E]%
\ColumnHook
\end{hscode}\resethooks

Here we observe the same restrictions as before, for example, if \ensuremath{\Varid{f}} yields \ensuremath{\Varid{failed}} for any argument \ensuremath{\Varid{x}} the equality does not hold.
Once again, we consider the context \ensuremath{\Varid{filterDist}\;\Varid{pred}} for all \ensuremath{\Varid{pred}\mathbin{::}\tau _{2}\to \Conid{Bool}} to reason that the user cannot observe the difference.

\begin{hscode}\SaveRestoreHook
\column{B}{@{}>{\hspre}l<{\hspost}@{}}%
\column{E}{@{}>{\hspre}l<{\hspost}@{}}%
\>[B]{}\Varid{filterDist}\;\Varid{pred}\;(\mathbf{let}\;\Conid{Dist}\;\Varid{z}\;\Varid{q}\mathrel{=}\Varid{f}\;\Varid{x}\;\mathbf{in}\;\Conid{Dist}\;\Varid{z}\;\Varid{q}){}\<[E]%
\\
\>[B]{}\equiv \mbox{\commentbegin  Definition of \ensuremath{\Varid{filterDist}}  \commentend}{}\<[E]%
\\
\>[B]{}\mathbf{let}\;\Conid{Dist}\;\Varid{x}\;\Varid{p}\mathrel{=}(\mathbf{let}\;\Conid{Dist}\;\Varid{z}\;\Varid{q}\mathrel{=}\Varid{f}\;\Varid{x}\;\mathbf{in}\;\Conid{Dist}\;\Varid{z}\;\Varid{q}){}\<[E]%
\\
\>[B]{}\mathbf{in}\;\mathbf{if}\;(\Varid{pred}\;\Varid{x})\;\mathbf{then}\;(\Conid{Dist}\;\Varid{x}\;\Varid{p})\;\mathbf{else}\;\Varid{failed}{}\<[E]%
\\
\>[B]{}\equiv \mbox{\commentbegin  Inline \ensuremath{\mathbf{let}}-declaration  \commentend}{}\<[E]%
\\
\>[B]{}\mathbf{let}\;\Conid{Dist}\;\Varid{x}\;\Varid{p}\mathrel{=}\Varid{f}\;\Varid{x}{}\<[E]%
\\
\>[B]{}\mathbf{in}\;\mathbf{if}\;(\Varid{pred}\;\Varid{x})\;\mathbf{then}\;(\Conid{Dist}\;\Varid{x}\;\Varid{p})\;\mathbf{else}\;\Varid{failed}{}\<[E]%
\\
\>[B]{}\equiv \mbox{\commentbegin  Definition of \ensuremath{\Varid{filterDist}}  \commentend}{}\<[E]%
\\
\>[B]{}\Varid{filterDist}\;\Varid{pred}\;(\Varid{f}\;\Varid{x}){}\<[E]%
\ColumnHook
\end{hscode}\resethooks

Fortunately, the second monad law holds as well in the context of \ensuremath{\Varid{filterDist}}.

\paragraph{The third monad law}
We do not discuss the third monad law \ref{item:law3} in detail here, as it holds without restrictions --- even without the additional context using \ensuremath{\Varid{filterDist}}.
All in all, \ensuremath{\Varid{certainly}} and \ensuremath{(\mathbin{>\!\!\!>\!\!\!>\!=})} form a valid monad from the user's point of view.

\section{Applications and Evaluation}
\label{sec:applications}

After presenting the basic combinators of the library and motivating
the advantages of modeling distributions using non-determinism, we
will implement some exemplary applications.
We reimplement examples that have been characterized as challenging
for probabilistic logic programming by Nampally and Ramakrishnan \cite{nampally2015constraintbased}, who use the examples to discuss
the expressiveness of probabilistic logic programming and its cost
with respect to performance.
The examples focus on properties of random strings and their probabilities.
Furthermore, we show benchmarks of these examples and compare them with the probabilistic languages ProbLog \cite{kimmig2011implementation} and WebPPL \cite{goodman2014design}.
These comparisons confirm the advantages of non-strict non-determinism with respect to performance.

\subsection{Random Strings}

In order to compare our library with other approaches for
probabilistic programming we reimplement two examples about random
strings that have also been implemented in ProbLog.
This implementation can be found online.\footnote{\url{https://dtai.cs.kuleuven.be/problog/tutorial/various/04_nampally.html}}
We generate random strings of a fixed length over the alphabet $\{a,b\}$ and calculate the probability that this string a) is a palindrome and b) contains the subsequence $bb$.

First we define a distribution that picks a character uniformly from the alphabet $\{a,b\}$.
\begin{hscode}\SaveRestoreHook
\column{B}{@{}>{\hspre}l<{\hspost}@{}}%
\column{3}{@{}>{\hspre}l<{\hspost}@{}}%
\column{E}{@{}>{\hspre}l<{\hspost}@{}}%
\>[3]{}\Varid{pickChar}\mathbin{::}\Conid{Dist}\;\Conid{Char}{}\<[E]%
\\
\>[3]{}\Varid{pickChar}\mathrel{=}\Varid{uniform}\;[\mskip1.5mu \text{\tt 'a'},\text{\tt 'b'}\mskip1.5mu]{}\<[E]%
\ColumnHook
\end{hscode}\resethooks
Based on \ensuremath{\Varid{pickChar}} we define a distribution that generates a random string of length \ensuremath{\Varid{n}}, that is, picks a random char \ensuremath{\Varid{n}} times.
We reuse \ensuremath{\Varid{replicateDist}} to define this distribution.
\begin{hscode}\SaveRestoreHook
\column{B}{@{}>{\hspre}l<{\hspost}@{}}%
\column{3}{@{}>{\hspre}l<{\hspost}@{}}%
\column{E}{@{}>{\hspre}l<{\hspost}@{}}%
\>[3]{}\Varid{randomString}\mathbin{::}\Conid{Int}\to \Conid{Dist}\;\Conid{String}{}\<[E]%
\\
\>[3]{}\Varid{randomString}\;\Varid{n}\mathrel{=}\Varid{replicateDist}\;\Varid{n}\;(\lambda ()\to \Varid{pickChar}){}\<[E]%
\ColumnHook
\end{hscode}\resethooks

In order to compute the probability that a random string is a palindrome and contains a subsequence $bb$, respectively, we define predicates that test these properties for a given string.
A string is a palindrome, if it reads the same forwards and backwards.
The following predicate, thus, checks if the reverse of a given string is equal to the original string.

\begin{hscode}\SaveRestoreHook
\column{B}{@{}>{\hspre}l<{\hspost}@{}}%
\column{3}{@{}>{\hspre}l<{\hspost}@{}}%
\column{E}{@{}>{\hspre}l<{\hspost}@{}}%
\>[3]{}\Varid{palindrome}\mathbin{::}\Conid{String}\to \Conid{Bool}{}\<[E]%
\\
\>[3]{}\Varid{palindrome}\;\Varid{str}\mathrel{=}\Varid{str}\equiv \Varid{reverse}\;\Varid{str}{}\<[E]%
\ColumnHook
\end{hscode}\resethooks
The predicate that checks if a string contains two consecutive $b$s can be easily defined via pattern matching and recursion.
\begin{hscode}\SaveRestoreHook
\column{B}{@{}>{\hspre}l<{\hspost}@{}}%
\column{3}{@{}>{\hspre}l<{\hspost}@{}}%
\column{29}{@{}>{\hspre}l<{\hspost}@{}}%
\column{50}{@{}>{\hspre}l<{\hspost}@{}}%
\column{E}{@{}>{\hspre}l<{\hspost}@{}}%
\>[3]{}\Varid{consecutiveBs}\mathbin{::}\Conid{String}\to \Conid{Bool}{}\<[E]%
\\
\>[3]{}\Varid{consecutiveBs}\;\Varid{str}\mathrel{=}\mathbf{case}\;{}\<[29]%
\>[29]{}\Varid{str}\;\mathbf{of}{}\<[E]%
\\
\>[29]{}[\mskip1.5mu \mskip1.5mu]{}\<[50]%
\>[50]{}\to \Conid{False}{}\<[E]%
\\
\>[29]{}(\text{\tt 'b'}\mathbin{:}\text{\tt 'b'}\mathbin{:}\Varid{rest}){}\<[50]%
\>[50]{}\to \Conid{True}{}\<[E]%
\\
\>[29]{}(\Varid{c}\mathbin{:}\Varid{rest}){}\<[50]%
\>[50]{}\to \Varid{consecutiveBs}\;\Varid{rest}{}\<[E]%
\ColumnHook
\end{hscode}\resethooks

Now we are ready to perform some queries.
What is the probability that a random string of length \ensuremath{\mathrm{5}} is a palindrome?

\begin{hscode}\SaveRestoreHook
\column{B}{@{}>{\hspre}l<{\hspost}@{}}%
\column{E}{@{}>{\hspre}l<{\hspost}@{}}%
\>[B]{}\mkern-3mu\mathbin{>}\Varid{palindrome}\mathbin{??}(\Varid{randomString}\;\mathrm{5}){}\<[E]%
\\
\>[B]{}\mathrm{0.25}{}\<[E]%
\ColumnHook
\end{hscode}\resethooks

\noindent What is the probability that a random string of length \ensuremath{\mathrm{10}} contains two consecutive $b$s?
\begin{hscode}\SaveRestoreHook
\column{B}{@{}>{\hspre}l<{\hspost}@{}}%
\column{E}{@{}>{\hspre}l<{\hspost}@{}}%
\>[B]{}\mkern-3mu\mathbin{>}\Varid{consecutiveBs}\mathbin{??}(\Varid{randomString}\;\mathrm{10}){}\<[E]%
\\
\>[B]{}\mathrm{0.859375}{}\<[E]%
\ColumnHook
\end{hscode}\resethooks

In general the above definitions of \ensuremath{\Varid{palindrome}} and \ensuremath{\Varid{consecutiveBs}} are quite naive and, thus, inefficient because all strings of the given length have to be enumerated explicitly.
Due to the inefficiency, the ProbLog homepage introduces a more efficient version for both problems.
In the following, we will discuss the alternative implementation to compute the probability for a palindrome only.
This more efficient version has arguments for the index of the front and back position, picks characters for both ends and then moves the position towards the middle.
That is, instead of naively generating the whole string of length $n$, this version checks each pair of front and back position first and fails straightaway, if they do not match.
If the characters do match, the approach continues by moving both
indices towards each other.
In Curry an implementation of this idea looks as follows.
\begin{hscode}\SaveRestoreHook
\column{B}{@{}>{\hspre}l<{\hspost}@{}}%
\column{3}{@{}>{\hspre}l<{\hspost}@{}}%
\column{22}{@{}>{\hspre}l<{\hspost}@{}}%
\column{35}{@{}>{\hspre}l<{\hspost}@{}}%
\column{38}{@{}>{\hspre}l<{\hspost}@{}}%
\column{E}{@{}>{\hspre}l<{\hspost}@{}}%
\>[3]{}\Varid{palindromeEfficient}\mathbin{::}\Conid{Int}\to \Conid{Dist}\;(\Conid{Bool},\Conid{String}){}\<[E]%
\\
\>[3]{}\Varid{palindromeEfficient}\;\Varid{n}\mathrel{=}\Varid{palindrome'}\;\mathrm{1}\;\Varid{n}{}\<[E]%
\\[\blanklineskip]%
\>[3]{}\Varid{palindrome'}\mathbin{::}\Conid{Int}\to \Conid{Int}\to \Conid{Dist}\;(\Conid{Bool},\Conid{String}){}\<[E]%
\\
\>[3]{}\Varid{palindrome'}\;\Varid{n1}\;\Varid{n2}{}\<[22]%
\>[22]{}\mid \Varid{n1}\equiv \Varid{n2}{}\<[35]%
\>[35]{}\mathrel{=}\Varid{pickChar}\mathbin{>\!\!\!>\!\!\!>\!=}(\lambda \Varid{c}\to \Varid{certainly}\;(\Conid{True},[\mskip1.5mu \Varid{c}\mskip1.5mu])){}\<[E]%
\\
\>[22]{}\mid \Varid{n1}\mathbin{>}\Varid{n2}{}\<[35]%
\>[35]{}\mathrel{=}\Varid{certainly}\;(\Conid{True},[\mskip1.5mu \mskip1.5mu]){}\<[E]%
\\
\>[22]{}\mid \Varid{otherwise}{}\<[35]%
\>[35]{}\mathrel{=}{}\<[38]%
\>[38]{}\Varid{pickChar}\mathbin{>\!\!\!>\!\!\!>\!=}(\lambda \Varid{c1}\to {}\<[E]%
\\
\>[38]{}\Varid{pickChar}\mathbin{>\!\!\!>\!\!\!>\!=}(\lambda \Varid{c2}\to {}\<[E]%
\\
\>[38]{}(\Varid{palindrome'}\;(\Varid{n1}\mathbin{+}\mathrm{1})\;(\Varid{n2}\mathbin{-}\mathrm{1}))\mathbin{>\!\!\!>\!\!\!>\!=}(\lambda (\Varid{b},\Varid{cs})\to {}\<[E]%
\\
\>[38]{}\Varid{certainly}\;(\Varid{c1}\equiv \Varid{c2}\mathrel{\wedge}\Varid{b},\Varid{c1}\mathbin{:}(\Varid{cs}\plus [\mskip1.5mu \Varid{c2}\mskip1.5mu]))))){}\<[E]%
\ColumnHook
\end{hscode}\resethooks

The interesting insight here is that, thanks to the combination of
non-determinism and non-strictness, the evaluation of the first query
based on \ensuremath{\Varid{palindrome}} behaves similar to the efficient variant in ProbLog.
At first, it seems that the query performs poorly, because the predicate \ensuremath{\Varid{palindrome}} needs to evaluate the whole list due to the usage of \ensuremath{\Varid{reverse}}.
The good news is, however, that the non-determinism is only spawned if we evaluate the elements of that list, and the elements still evaluate non-strictly, when explicitly triggered by \ensuremath{(\equiv )}.
More precisely, because of the combination of \ensuremath{\Varid{reverse}} and \ensuremath{(\equiv )}, the evaluation starts by checking the first and last characters of a string and only continues to check more characters, and spawn more non-determinism, if they match.
If these characters do not match, the evaluation fails directly and does not need to check any more characters.
In a nutshell, when using PFLP, we get a version competitive with efficient implementations although
we used a naive generate and test approach.

\subsection{Performance Comparisons with Other Languages}
\label{ssec:performance}

Up to now, the only performance comparisons we discussed were for
different implementations of our library in Curry and Haskell.
These comparisons showed the advantage of using non-strict
non-determinism concepts for the implementation of the library.
Next we want to take a look at the comparison with the full-blown
probabilistic programming languages ProbLog and WebPPL.
ProbLog is a probabilistic extension of Prolog that is implemented in
Python.
WebPPL is the successor of Church; in contrast to Church it is not
implemented in Scheme but in JavaScript.

In order to try to measure the execution of the programs only, we
precompiled the executable for the Curry programs.
As Python is an interpreted language, a similar preparation was not
available for ProbLog.
However, we used ProbLog as a library in order to call the Python\footnote{We use version 2.7.10 of Python.} interpreter directly.
ProbLog is mainly implemented in Python, which allows users to
import ProbLog as a Python package.\footnote{\url{https://dtai.cs.kuleuven.be/problog/tutorial/advanced/01_python_interface.html}}
For WebPPL, we used \textit{node.js}\footnote{We use version 8.12.0 of \textit{node.js}.} to run the JavaScript program as a
terminal application.
All of the following running times are the mean of $1000$ runs
as calculated by the Haskell tool
bench\footnote{\url{https://hackage.haskell.org/package/bench}} that
we use to run the benchmarks.

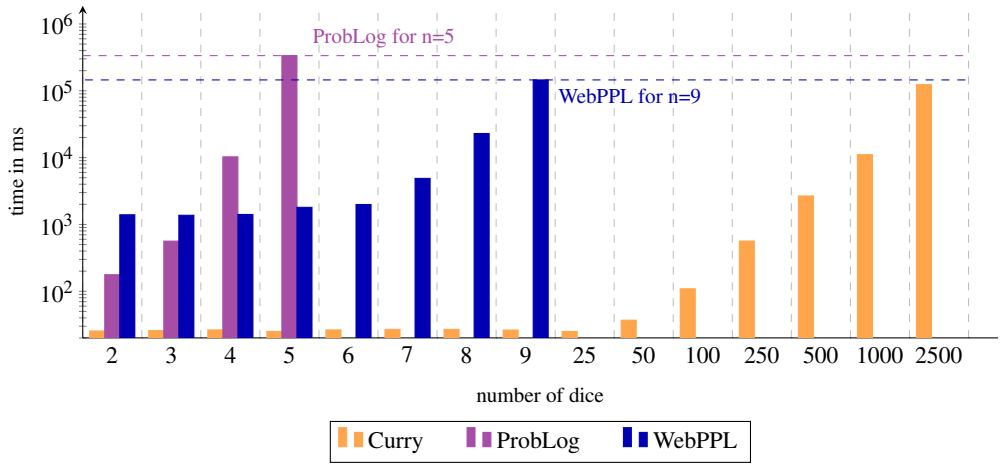
\begin{figure}
\input{figures/diceBarAll}
\caption{Getting only sixes when rolling $n$ dice}
\label{fig:dice}
 \end{figure}

We compare the running times based on the two examples we already
discussed: the dice rolling example presented in \autoref{ssec:nonstrict} and the palindrome example from the previous subsection.
\paragraph{Dice Rolling}
As discussed before, non-strict non-determinism performs pretty
well for the dice rolling example, as a great deal of the search space
is pruned early.
\autoref{fig:dice} shows an impressive advantage of our library in comparison with ProbLog and WebPPL.
The x-axis represents the number of rolled dice and we present
the time in milliseconds in logarithmic scale on the y-axis.

In order to demonstrate that our library outperforms ProbLog and WebPPL
by several orders of magnitude for this example, we also run the Curry
implementation for bigger values of $n$ that eventually had the same
running time as the last tested value for the other languages.
The right part of \autoref{fig:dice} shows the running times for $25$ to $5000$ dice.
We can see that our library can compute the probability for
getting only sixes for $2500$ dice in roughly the same time as ProbLog
for $5$ dice.
The running times for WebPPL seem very bad in the beginning, but
after a few throws it becomes obvious that there is a constant overhead.
In fact, Nogatz et al. \cite{nogatz2018chr} observe and discuss this overhead as well.
Nevertheless, whereas WebPPL computes the probability for $9$ dice,
our library can compute the probability for $2500$ dice in roughly the same
time.

\paragraph{Palindrome}

\begin{figure}[ht]
\input{figures/palindromeBar}
\caption{Palindrome computation for a string of length $n$}
\label{fig:palindromeBar}
\end{figure}
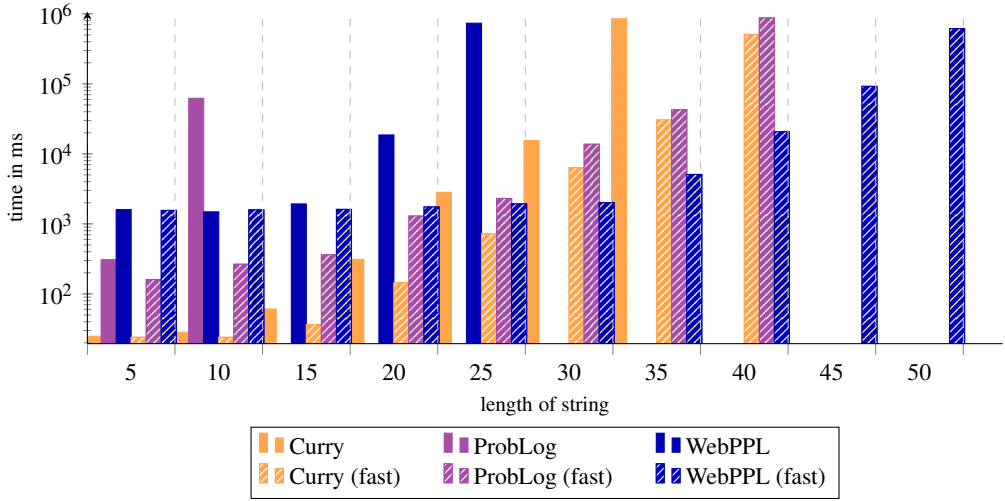

In order to back up the results of the previous example, \autoref{fig:palindromeBar} shows
benchmarks for implementations of the naive and the efficient versions
in Curry, ProbLog and WebPPL.
The x-axis represents the length of the generated palindrome and, once again,
we present the time in milliseconds in logarithmic scale on the y-axis.

The figure uses dashed bars for the efficient version of the algorithm and a solid filling for the naive algorithm.
The naive algorithm scales pretty bad in ProbLog and WebPPL.
The Curry version is still applicable up to $30$ as its running time is similar
to all three efficient versions.
Overall, the efficient versions all perform in a similar time range, but
WebPPL shows a slight performance advantage for an increasing
length of the string.
More precisely, the efficient WebPPL implementation performs a query for strings of length $50$ in the same time as the efficient Curry and ProbLog perform a query for strings of length $40$.
That is, the efficient WebPPL implementation outperforms the other implementations by roughly two orders of magnitude.

\section{Related and Future Work}

The approach of this paper is based on the work by Erwig and Kollmansberger \cite{erwig2006functional}, who introduce a Haskell library that represents distributions as lists of event-probability pairs. %
Their library also provides a simple sampling mechanism to perform inference on distributions.
Inference algorithms come into play because common examples in probabilistic programming have an exponential growth and it is not feasible to compute the whole distribution.
Similarly, \'{S}cibior et al. \cite{scibior2015practical} present a more efficient implementation using a DSL in Haskell.
They represent distributions as a free monad and inference algorithms as an interpretation of the monadic structure.
Thanks to this interpretation, the approach is competitive to full-blown probabilistic programming languages with respect to performance.
PFLP provides functions to sample from distributions as well.
However, in this work we focus on modeling distributions and do not discuss any sampling mechanism.
In particular, as future work we plan to investigate whether we can benefit from the improved performance as presented here in the case of sampling.
Furthermore, a more detailed investigation of the performance of non-determinism in comparison to a list model is a topic for another paper.

The benefit with respect to the combination of non-strictness and non-de\-ter\-min\-ism is similar to the benefit of property-based testing using Curry-like non-determinism in Haskell \cite{runciman2008smallcheck} and Curry \cite{christiansen2008easycheck}.
In property-based testing, sometimes we want to generate test cases that satisfy a precondition.
With Curry-like non-determinism the precondition can prune the search space early, while a list-based implementation has to generate all test cases and filter them afterwards.
Both applications, probabilistic programming and property-based testing, are examples, where built-in non-determinism outperforms list-based approaches as introduced by Wadler \cite{wadler1985replace}.
In comparison to property-based testing, here, we observe that we can even add a kind of monadic layer on top of the non-determinism that computes additional information and still preserve the demand-driven behavior.
However, the additional information has to be evaluated strictly --- as it is the case for probabilities, otherwise we might lose non-deterministic results.

There are other more elaborated approaches to implement a library for probabilistic programming.
For example, Kiselyov and Shan \cite{kiselyov2009embedded} extend their library for probabilistic programming in OCaml with a construct for lazy evaluation to achieve similar behavior with respect to efficiency.
However, they use lazy evaluation for a concrete application based on importance sampling.
Due to the combination of non-strictness and non-determinism, we can efficiently calculate the total probability of the resulting distribution without utilizing sampling.

As future work, we see a high potential for performance improvements for the Curry compiler KiCS2.
PFLP serves as a starting point for further studies of functional logic features in practical applications.
For example, we would expect the running times of the strict implementation based on non-determinism to be approximately as efficient as a list-based implementation.
However, as the numbers in \autoref{sec:details} show, the list approach is considerably faster.

The library's design does not support the use of non-determinism in events or probabilities of a distribution.
In case of deeper non-determinism, we have to be careful to trigger all non-determinism when querying a distribution as shown in \autoref{sec:details}.
Hence, the extension of the library with an interface using non-determinism on the user's side is an idea worth studying.

Last but not least, we see an opportunity to apply ideas and solutions of the functional logic paradigm in probabilistic programming.
For instance,~Christiansen et al. \cite{christiansen2010free} investigate free theorems for functional logic programs.
As their work considers non-determinism and sharing, adapting it to probabilistic programming should be easy.
As another example,~Bra\ss{}el \cite{brassel2009technique} presents a debugger for Curry that works well with non-determinism.
Hence, it should be possible to reuse these ideas in the setting of probabilistic programming as well.

\section{Conclusion}

We have implemented a simple library for probabilistic programming in a functional logic programming language, namely Curry.
Such a library proves to be a good fit for a functional logic language, because both paradigms share similar features.
While other libraries need to reimplement features specific to probabilistic programming, we solely rely on core features of functional logic languages.

The key idea of the library is to use non-determinism to model distributions.
We discussed design choices as well as the disadvantages and advantages that result from this approach.
In the end, the library provides non-strict probabilistic combinators in order to avoid spawning unnecessary non-deterministic computations.
These non-strict combinators have benefits in terms of performance due to early pruning.
Using combinators that are too strict leads to a loss of these performance benefits.
Fortunately, the user does not have to worry about using the right amount of strictness as long as she only uses the provided combinators.

There are, however, two restrictions the user has to follow when using the library.
If the user does not follow these restrictions, a program may behave unexpectedly.
Events may not be non-deterministic and the second argument of \ensuremath{(\mathbin{>\!\!\!>\!\!\!>\!=})}-operator may not be partial.
Notwithstanding, we want to emphasize that the restrictions do not affect expressibility.
In fact, a programming language like ProbLog shows similar behavior when mixing non-determinism and probabilities as our implementation.

Last but not least, we showed that the library obeys the expected monad laws with respect to observational behavior and reimplemented examples from the probabilistic programming literature to compare the performance of our library
with other existing languages.

\bigskip
\noindent
\textbf{Acknowledgements}~
We are thankful for fruitful discussions with Michael Hanus as well as suggestions of Jan Bracker and Falco Nogatz.
Finally, we are thankful for the comments of the anonymous reviewers to improve the readability of this paper.

\bibliography{tplp}

\appendix{}

\section{Predefined Curry Functions}
\label{appendix:functions}

This section presents all predefined functions that are used in this paper.
\begin{hscode}\SaveRestoreHook
\column{B}{@{}>{\hspre}l<{\hspost}@{}}%
\column{8}{@{}>{\hspre}c<{\hspost}@{}}%
\column{8E}{@{}l@{}}%
\column{12}{@{}>{\hspre}l<{\hspost}@{}}%
\column{15}{@{}>{\hspre}l<{\hspost}@{}}%
\column{E}{@{}>{\hspre}l<{\hspost}@{}}%
\>[B]{}(\mathrel{\wedge})\mathbin{::}\Conid{Bool}\to \Conid{Bool}\to \Conid{Bool}{}\<[E]%
\\
\>[B]{}\Conid{False}{}\<[8]%
\>[8]{}\mathrel{\wedge}{}\<[8E]%
\>[12]{}\Varid{x}{}\<[15]%
\>[15]{}\mathrel{=}\Conid{False}{}\<[E]%
\\
\>[B]{}\Conid{True}{}\<[8]%
\>[8]{}\mathrel{\wedge}{}\<[8E]%
\>[12]{}\Varid{x}{}\<[15]%
\>[15]{}\mathrel{=}\Varid{x}{}\<[E]%
\ColumnHook
\end{hscode}\resethooks
\begin{hscode}\SaveRestoreHook
\column{B}{@{}>{\hspre}l<{\hspost}@{}}%
\column{8}{@{}>{\hspre}l<{\hspost}@{}}%
\column{11}{@{}>{\hspre}l<{\hspost}@{}}%
\column{14}{@{}>{\hspre}l<{\hspost}@{}}%
\column{22}{@{}>{\hspre}l<{\hspost}@{}}%
\column{E}{@{}>{\hspre}l<{\hspost}@{}}%
\>[B]{}\Varid{foldr}\mathbin{::}(\Varid{a}\to \Varid{b}\to \Varid{b})\to \Varid{b}\to [\mskip1.5mu \Varid{a}\mskip1.5mu]\to \Varid{b}{}\<[E]%
\\
\>[B]{}\Varid{foldr}\;{}\<[8]%
\>[8]{}\Varid{f}\;{}\<[11]%
\>[11]{}\Varid{z}\;{}\<[14]%
\>[14]{}[\mskip1.5mu \mskip1.5mu]{}\<[22]%
\>[22]{}\mathrel{=}\Varid{z}{}\<[E]%
\\
\>[B]{}\Varid{foldr}\;{}\<[8]%
\>[8]{}\Varid{f}\;{}\<[11]%
\>[11]{}\Varid{z}\;{}\<[14]%
\>[14]{}(\Varid{x}\mathbin{:}\Varid{xs}){}\<[22]%
\>[22]{}\mathrel{=}\Varid{f}\;\Varid{x}\;(\Varid{foldr}\;\Varid{f}\;\Varid{z}\;\Varid{xs}){}\<[E]%
\ColumnHook
\end{hscode}\resethooks
\begin{hscode}\SaveRestoreHook
\column{B}{@{}>{\hspre}l<{\hspost}@{}}%
\column{10}{@{}>{\hspre}l<{\hspost}@{}}%
\column{13}{@{}>{\hspre}l<{\hspost}@{}}%
\column{23}{@{}>{\hspre}l<{\hspost}@{}}%
\column{33}{@{}>{\hspre}l<{\hspost}@{}}%
\column{E}{@{}>{\hspre}l<{\hspost}@{}}%
\>[B]{}\Varid{zipWith}\mathbin{::}(\Varid{a}\to \Varid{b}\to \Varid{c})\to [\mskip1.5mu \Varid{a}\mskip1.5mu]\to [\mskip1.5mu \Varid{b}\mskip1.5mu]\to [\mskip1.5mu \Varid{c}\mskip1.5mu]{}\<[E]%
\\
\>[B]{}\Varid{zipWith}\;{}\<[10]%
\>[10]{}\Varid{f}\;{}\<[13]%
\>[13]{}[\mskip1.5mu \mskip1.5mu]\;{}\<[23]%
\>[23]{}\Varid{ys}{}\<[33]%
\>[33]{}\mathrel{=}[\mskip1.5mu \mskip1.5mu]{}\<[E]%
\\
\>[B]{}\Varid{zipWith}\;{}\<[10]%
\>[10]{}\Varid{f}\;{}\<[13]%
\>[13]{}(\Varid{x}\mathbin{:}\Varid{xs})\;{}\<[23]%
\>[23]{}[\mskip1.5mu \mskip1.5mu]{}\<[33]%
\>[33]{}\mathrel{=}[\mskip1.5mu \mskip1.5mu]{}\<[E]%
\\
\>[B]{}\Varid{zipWith}\;{}\<[10]%
\>[10]{}\Varid{f}\;{}\<[13]%
\>[13]{}(\Varid{x}\mathbin{:}\Varid{xs})\;{}\<[23]%
\>[23]{}(\Varid{y}\mathbin{:}\Varid{ys}){}\<[33]%
\>[33]{}\mathrel{=}\Varid{f}\;\Varid{x}\;\Varid{y}\mathbin{:}\Varid{zipWith}\;\Varid{f}\;\Varid{xs}\;\Varid{ys}{}\<[E]%
\ColumnHook
\end{hscode}\resethooks
\begin{hscode}\SaveRestoreHook
\column{B}{@{}>{\hspre}l<{\hspost}@{}}%
\column{E}{@{}>{\hspre}l<{\hspost}@{}}%
\>[B]{}\Varid{repeat}\mathbin{::}\Varid{a}\to [\mskip1.5mu \Varid{a}\mskip1.5mu]{}\<[E]%
\\
\>[B]{}\Varid{repeat}\;\Varid{x}\mathrel{=}\Varid{x}\mathbin{:}\Varid{repeat}\;\Varid{x}{}\<[E]%
\ColumnHook
\end{hscode}\resethooks
\begin{hscode}\SaveRestoreHook
\column{B}{@{}>{\hspre}l<{\hspost}@{}}%
\column{18}{@{}>{\hspre}l<{\hspost}@{}}%
\column{E}{@{}>{\hspre}l<{\hspost}@{}}%
\>[B]{}\Varid{length}\mathbin{::}[\mskip1.5mu \Varid{a}\mskip1.5mu]\to \Conid{Int}{}\<[E]%
\\
\>[B]{}\Varid{length}\;[\mskip1.5mu \mskip1.5mu]{}\<[18]%
\>[18]{}\mathrel{=}\mathrm{0}{}\<[E]%
\\
\>[B]{}\Varid{length}\;(\Varid{x}\mathbin{:}\Varid{xs}){}\<[18]%
\>[18]{}\mathrel{=}\mathrm{1}\mathbin{+}\Varid{length}\;\Varid{xs}{}\<[E]%
\ColumnHook
\end{hscode}\resethooks
\begin{hscode}\SaveRestoreHook
\column{B}{@{}>{\hspre}l<{\hspost}@{}}%
\column{9}{@{}>{\hspre}l<{\hspost}@{}}%
\column{12}{@{}>{\hspre}l<{\hspost}@{}}%
\column{22}{@{}>{\hspre}l<{\hspost}@{}}%
\column{32}{@{}>{\hspre}l<{\hspost}@{}}%
\column{38}{@{}>{\hspre}l<{\hspost}@{}}%
\column{55}{@{}>{\hspre}l<{\hspost}@{}}%
\column{61}{@{}>{\hspre}l<{\hspost}@{}}%
\column{E}{@{}>{\hspre}l<{\hspost}@{}}%
\>[B]{}\Varid{filter}\mathbin{::}(\Varid{a}\to \Conid{Bool})\to [\mskip1.5mu \Varid{a}\mskip1.5mu]\to [\mskip1.5mu \Varid{a}\mskip1.5mu]{}\<[E]%
\\
\>[B]{}\Varid{filter}\;{}\<[9]%
\>[9]{}\Varid{p}\;{}\<[12]%
\>[12]{}[\mskip1.5mu \mskip1.5mu]{}\<[22]%
\>[22]{}\mathrel{=}[\mskip1.5mu \mskip1.5mu]{}\<[E]%
\\
\>[B]{}\Varid{filter}\;{}\<[9]%
\>[9]{}\Varid{p}\;{}\<[12]%
\>[12]{}(\Varid{x}\mathbin{:}\Varid{xs}){}\<[22]%
\>[22]{}\mathrel{=}\mathbf{if}\;\Varid{p}\;\Varid{x}\;{}\<[32]%
\>[32]{}\mathbf{then}\;{}\<[38]%
\>[38]{}\Varid{x}\mathbin{:}\Varid{filter}\;\Varid{p}\;\Varid{xs}\;{}\<[55]%
\>[55]{}\mathbf{else}\;{}\<[61]%
\>[61]{}\Varid{filter}\;\Varid{p}\;\Varid{xs}{}\<[E]%
\ColumnHook
\end{hscode}\resethooks
\begin{hscode}\SaveRestoreHook
\column{B}{@{}>{\hspre}l<{\hspost}@{}}%
\column{E}{@{}>{\hspre}l<{\hspost}@{}}%
\>[B]{}\Varid{all}\mathbin{::}(\Varid{a}\to \Conid{Bool})\to [\mskip1.5mu \Varid{a}\mskip1.5mu]\to \Conid{Bool}{}\<[E]%
\\
\>[B]{}\Varid{all}\;\Varid{p}\;\Varid{xs}\mathrel{=}\Varid{foldr}\;(\mathrel{\wedge})\;\Conid{True}\;(\Varid{map}\;\Varid{p}\;\Varid{xs}){}\<[E]%
\ColumnHook
\end{hscode}\resethooks
\begin{hscode}\SaveRestoreHook
\column{B}{@{}>{\hspre}l<{\hspost}@{}}%
\column{11}{@{}>{\hspre}l<{\hspost}@{}}%
\column{E}{@{}>{\hspre}l<{\hspost}@{}}%
\>[B]{}not\mathbin{::}\Conid{Bool}\to \Conid{Bool}{}\<[E]%
\\
\>[B]{}not\;\Conid{False}\mathrel{=}\Conid{True}{}\<[E]%
\\
\>[B]{}not\;\Conid{True}{}\<[11]%
\>[11]{}\mathrel{=}\Conid{False}{}\<[E]%
\ColumnHook
\end{hscode}\resethooks
\begin{hscode}\SaveRestoreHook
\column{B}{@{}>{\hspre}l<{\hspost}@{}}%
\column{E}{@{}>{\hspre}l<{\hspost}@{}}%
\>[B]{}\Varid{const}\mathbin{::}\Varid{a}\to \Varid{b}\to \Varid{a}{}\<[E]%
\\
\>[B]{}\Varid{const}\;\Varid{x}\;\Varid{y}\mathrel{=}\Varid{x}{}\<[E]%
\ColumnHook
\end{hscode}\resethooks

\end{document}

%% file: figures/diceBarAll.tex
\begin{tikzpicture}
  \begin{semilogyaxis}[
    tickwidth = 0pt,
    xmajorgrids=true,
    grid style={dashed},
    height=6cm,
    width=13.75cm,
    bar width=0.2cm,
    ybar=0pt,
    xlabel = {\footnotesize{number of dice}},
    ylabel = {\footnotesize{time in ms}},
    ymax = 1500000,
    xmax = 16,
    xmin = 0.5,
    axis y line = left,
    xtick = {0.5,1.5,2.5,...,16.5},
    xticklabels = {2,3,4,5,6,7,8,9,25,50,100,250,500,1000,2500,5000},
    x tick label as interval,
    axis x line* = bottom,
    xlabel style={at={(0.5,-0.125)}},
    enlarge y limits  = 0.02,
    legend style={
      at={(0.5,-0.25)},
      anchor=north,
      legend columns=-1,
      /tikz/every even column/.append style={column sep=0.5cm}}
  ]

  \addplot+[curry]
    coordinates {
                (1,25.48)
                (2,25.83)
                (3,26.42)
                (4,25.11)
                (5,26.43)
                (6,26.91)
                (7,26.88)
                (8,26.31)
                (9,25.01)
                (10,36.97)
                (11,109.0)
                (12,564.6)
                (13,2668)
                (14,11080)
                (15,123500)
    };

  \addplot+[probLog]
    coordinates {
                (1,177.1)
                (2,563.1)
                (3,10270)
                (4,335700)
    };

  \addplot+[webPPL]
    coordinates {
                (1,1396)
                (2,1373)
                (3,1411)
                (4,1805)
                (5,1989)
                (6,4890)
                (7,22960)
                (8,145400)
    };

 \draw [probLog,dashed] (0.5, 335700)--(15.5, 335700)
    node[pos = 0.34,above] {\footnotesize{ProbLog for n=5}};

 \draw [webPPL,dashed] (0, 145400)--(15.5, 145400)
   node[pos = 0.63,below] {\footnotesize{WebPPL for n=9}};

  \legend{Curry, ProbLog, WebPPL}
  \end{semilogyaxis}

\end{tikzpicture}

%% file: figures/palindromeBar.tex
\begin{tikzpicture}
  \begin{semilogyaxis}[
    height=6cm,
    width=13.75cm,
    ybar=0pt,
    bar width=0.2cm,
    axis y line = left,
    xlabel = {\footnotesize{length of string}},
    ylabel = {\footnotesize{time in ms}},
    xmajorgrids=true,
    grid style={dashed},
    xtick = {2.5,7.5,...,55},
    xticklabels = {5,10,...,50},
    xmin=2.5,
    x tick label as interval,
    axis x line* = bottom,
    enlarge y limits  = 0.02,
    axis on top,
    legend cell align = {left},
    legend style={
      at={(0.5,-0.25)},
      anchor=north,
      legend columns=3,
      cells = {align = left},
      /tikz/every even column/.append style={column sep=0.5cm}}
  ]

  ]
  \addplot+[curry]
    coordinates {
      (35,851000)
      (30,15530)
      (25,2819)
      (20,310.7)
      (15,60.96)
      (10,28.33)
      (5,24.81)
    };

  \addplot+[probLog]
    coordinates {
      (10,62350)
      (5,307.9)
    };

  \addplot+[webPPL]
    coordinates {
      (25,735600)
      (20,18630)
      (15,1928)
      (10,1486)
      (5,1598) 
    };

  \addplot+[curry,  postaction={ pattern=north east
    lines, pattern color = white}]
    coordinates {
      (40,508500)
      (35,30510)
      (30,6342)
      (25,720.8)
      (20,144.3)  
      (15,36.37)
      (10,23.99)
      (5,24)
    };

  \addplot+[probLog,  postaction={ pattern=north east
    lines, pattern color = white}]
    coordinates {
      (40,874000)
      (35,42880)
      (30,13730)
      (25,2309)
      (20,1293)
      (15,361.3)
      (10,265.1)
      (5,159.8)
    };

  \addplot+[webPPL,  postaction={ pattern=north east
    lines, pattern color = white}]
    coordinates {
      (50,616000)
      (45,92140)
      (40,20660)
      (35,5081)
      (30,2004)
      (25,1930)
      (20,1745)
      (15,1610)
      (10,1586)
      (5,1556)
    };

  \legend{Curry, ProbLog, WebPPL, Curry (fast), ProbLog (fast), WebPPL (fast)}
  \end{semilogyaxis}
\end{tikzpicture}